# General Theory of Absorption in Porous Materials: The Restricted Multilayer Theory

Alexander A. Aduenko,[†] Andy Murray,[‡] and Jose L. Mendoza-Cortes[*,‡,¶,§]

[†]Department of Control and Applied Mathematics, Moscow Institute of Physics and Technology, Dolgoprudny, 141700, Russia

[‡]Department of Scientific Computing, Materials Science and Engineering Program, High Performance Materials Institute, Condensed Matter Theory - National High Magnetic Field Laboratory, Florida State University, Tallahassee FL, 32310, USA

[¶]Department of Chemical & Biomedical Engineering, FAMU-FSU Joint College of Engineering, Tallahassee FL, 32310, USA

[§]Department of Physics, College of Arts and Science, Florida State University, Tallahassee FL, 32310, USA.

E-mail: mendoza@eng.famu.fsu.edu

Phone: +1-850-410-6298. Fax: +1-850-410-6150

## Abstract

In this paper we present an approach for the generalization of adsorption of light gases in porous materials. This new theory goes beyond the Langmuir and BET theories, which are the standard approaches that have limited application to crystalline porous materials by their unphysical assumptions on the amount of possible adsorption layers. The derivation of the more general equation for any crystalline porous framework is presented; the Restricted Multilayer Theory (RMT). Our approach allows the determination of gas uptake considering only geometrical constraints of the porous framework and the interaction energy of the guest molecule with the framework. Based on this theory, we calculated optimal values for the adsorption enthalpy at different temperatures and pressures. We also present the use of this theory to determine the optimal linker length for a topological equivalent framework series. We validate this theoretical approach by applying it to Metal-Organic Frameworks (MOF) and show that it reproduces the experimental results for 7 different reported materials. We obtained the universal equation for optimal linker length given the topology of a porous framework. This work applied the general equation to MOFs and $H_2$ to create energy storage materials, however this theory can be applied to other crystalline porous materials and light gases, which opens the possibility of designing the next generations of energy storage materials by first considering only the geometrical constraints of the porous materials.

# Keywords

$H_2$ storage, Metal-Organic framework (MOF), crystalline porous material, optimal enthalpy, delivery amount, general theory of absorption.

# 1 Introduction

Currently there are no simple guidelines for the optimal linker length, pore size and storage capacity on crystalline porous frameworks. If such guidelines existed, they could be used to design new porous materials for efficient energy storage of substances like molecular hydrogen $H_2$ or other molecules. The first set of theories available to estimate the adsorption properties of surfaces were the Langmuir theory, published in 1916, followed by the Brunauer-Emmett-Teller (BET) theory, published in 1938.[1,2] Though very useful, these two theories have limited application to crystalline porous materials by their unphysical assumptions on the amount of possible adsorption layers. Therefore, we present an approach that allows the determination of gas uptake in crystalline porous materials considering only their geometrical constraints and their interaction energy with the guest molecule. The derivation of this general equation for the uptake of crystalline porous frameworks was validated with experimentally available data for Metal-Organic Frameworks (MOF). We applied this general equation to calculate the net storage capacity of molecular hydrogen storage ($H_2$) in MOFs. In this work, we propose the ideal interaction energy and the ideal linker length for MOFs given a topology (or crystalline lattice as in Figure 1); however, in principle, this approach can be extended to any combination of gas with crystalline porous material.

This paper is organized into several sections. In the first section, we present the general equation for adsorption in any framework, given its geometric properties (i.e. void volume, pore volume, surface area and pore size) and enthalpy of adsorption with the guest molecule (i.e. interaction energy with the surface). Frameworks with several types of topologies (geometrical shapes) are also considered. In the second section, we compare our predictions of adsorption isotherm to experimental measurements for system temperatures, $T$, equal to 77, 243, 273, and 298 K. In the final section, we present the optimal enthalpy of adsorption and optimal linker-length/void-volume giving the highest net storage capacity.

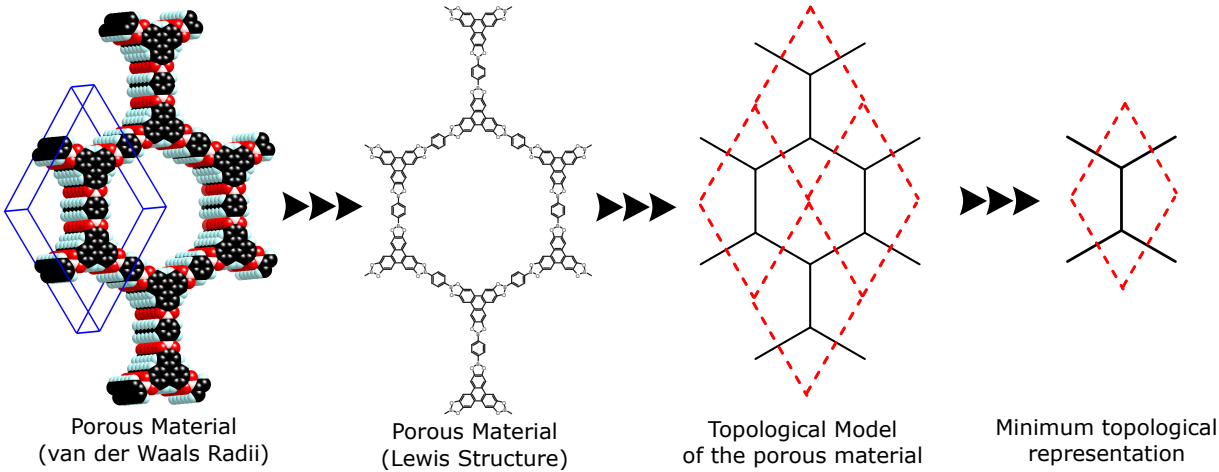

Porous Material
(van der Waals Radii)

Porous Material
(Lewis Structure)

Topological Model
of the porous material

Minimum topological
representation

Figure 1: A Porous Framework can be represented by a topological construction, i.e. in only edges and nodes. This principle can used for designing porous materials.

## 2　Methodology

**Restricted multilayer theory (RMT).**　The derivation of the general formula for the adsorption process is solely restricted to the topology of the framework. We define the topology of a framework as the connectivity in all directions to make a periodic structure. A simple example is shown in Figure 1. This topology is going to determine the number of adsorption sites available for our system and the limit to how many layers of adsorbate can fit.

The Langmuir theory assumes that only one layer of adsorbate molecules can be formed on the surface, whereas the BET theory assumes that an infinite number of layers can exist. Both assumptions are unphysical for a crystalline porous material, which can usually host more than one layer, but not an infinite number. For materials with small pores some gases cannot form even a single layer, hence having close to zero adsorption, which is not described by BET and Langmuir theories. In other words, both BET and Langmuir guarantee at least one adsorption layer which is not physical for gases with molecules bigger than the pore size. The BET theory also predicts infinite uptake for big enough pressure (See Supplementary Info Sec. IB), which is unphysical as well. Thus, a more appropriate theoretical approach is needed to treat the case of adsorption in crystalline porous materials, it is called restricted multilayer theory (RMT). The derivation of the Langmuir and BET theories can be found in the Suplementary Information and elsewhere.[1,2] We will show below that the equation for a restricted number of layers is a more general equation that produces the Langmuir and BET theories as particular solutions.

**Derivation of Restricted Multilayer Theory (RMT).**　For the general system (any combination of gas and porous materials) presented in Figure 2 we posit that there are $B$ equivalent sites, and $N$ molecules can be distributed in them. In Figure 2, we show examples of adsorbed sites represented by filled squares and empty squares representing the latter.

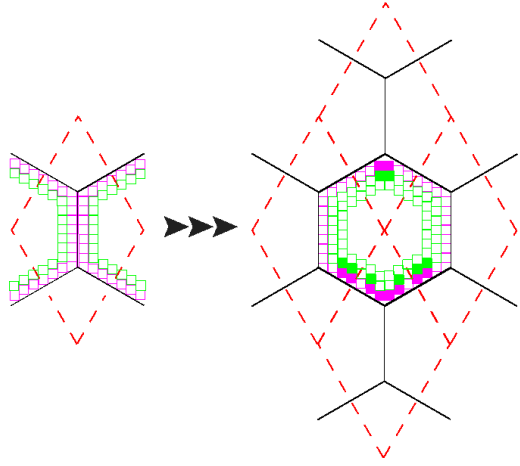

Figure 2: Restricted number of layers without (left) and with (right) adsorbed gas.

We assume that there are only three types of interactions in the system, which are

- Those between the surface ($n = 0$) and first layer molecules ($n = 1$), with partition function $j_1$;

- Those between first layer molecules ($n = 1$) and second layer molecules ($n = 2$), with partition function $j_2$, and

- Those between $n^{\text{th}}$ ($n > 1$) layer molecules and $n^{\text{th}} + 1$ layer molecules, with partition function $j_\infty$.

This creates a columnar structure of bound molecules. Additionally, we assume that there are no interactions between molecules of neighbouring columns and the partition function $j_2$ is equal to $j_\infty$. Given the total number of molecules, $N$, then $N_1$ of them are in the first layer and $N_n = N - N_1$ are in higher layers. From these set of assumptions, we can derive the adsorption behaviour for a restricted number of multilayers ($n$ layers).

Using the grand canonical partition function, we have

$$\Xi(B, \mu, T) = \sum_{N \geq 0} \exp[(N\mu)/kT] \, Q(N, B, T)$$

$$\sum_{N_1=0}^{B} \sum_{N_2=0}^{N_1} \sum_{N_3=0}^{N_2} \cdots \sum_{N_n=0}^{N_{n-1}} \frac{B!}{(B - N_1)! N_1!} \times \frac{N_1!}{(N_1 - N_2)! N_2!} \cdots \times \frac{N_{n-1}! (cx)^{N_1} x^{N_2 + \ldots + N_n}}{(N_{n-1} - N_n)! N_n!}. \tag{1}$$

In the preceding equations, $\mu$ is chemical potential, $T$ is temperature and $k$ is Boltzmann constant. The formula for chemical potential $\mu$ for molecular hydrogen both for low an high temperatures is derived in Supplementary Info (See Supplementary Info Sec. IB).

Summing in turn over $N_n$, $N_{n-1}$, ..., $N_1$ we find

$$\Xi = [1 + cx(1 + x + x^2 + \ldots + x^{n-1})]^B = \left[1 + cx\left(\frac{1 - x^n}{1 - x}\right)\right]^B \tag{2}$$

where we have defined,

$$c = j_1/j_\infty, \ \ x = j_\infty \exp(\mu/kT). \tag{3}$$

Using equation (3) in conjunction with the identity,

$$\overline{N} = kT\left(\frac{\partial \log \Xi}{\partial \mu}\right)_{T,B}, \tag{4}$$

we find the equation for the *adsorption of a restricted number of multilayers*:

$$\frac{\overline{N}}{B} = \frac{cx[1 - (n + 1)x^n + nx^{n+1}]}{(1 - x)(1 - x + cx - cx^{n+1})}. \tag{5}$$

This equation is more general because the substitution of $n = 1$ or $n = \infty$ gives the Langmuir and BET equations, respectively. Similar results were obtained earlier,[3,4] however in these earlier treatments; the topology of the framework, other geometrical constraints, and the specific gas properties were not taken into account. All these other variables are addressed in this work using the RMT approach.

**Application of RMT to $H_2$ in MOFs.** We first derive a general equation for the uptake, $m$, defined as the total mass of gas that can be incorporated per unit volume of a porous framework (in units of g/L). This equation applies for any framework based on the geometric constraints of the framework. Given that the molar mass of hydrogen $M = 2.018$ g/mol, the uptake per $V = 1\,L = 10^{27}\text{Å}^3$ equals

$$m = M\frac{N}{N_A} = M\frac{\gamma V}{V_p N_A}B\frac{N}{B}\left(\tilde{n}\right), \tag{6}$$

where $V_p$ is the volume of a pore, $B$ is the number of sites per pore, $N$ is a number of adsorbed hydrogen molecules per 1 L of adsorbent, $N_A$ is the Avogadro constant and $\tilde{n}$ is the effective number of layers estimated using geometrical properties of a pore (See Supplementary Info Sec. III). The factor $\gamma$ corresponds to the fraction of free (or void) volume for the framework and $V$ is the total volume, i.e. void volume $= \gamma V$. In Equation (6) we use that the number of pores per unit of volume is $\gamma V/V_p$. The full derivation can be found in the Supplementary Information Section IC. We also consider $D + \delta$ as the effective diameter of the adsorbed molecule which equals $2r_w = 2\cdot 2.76$ Å for a hydrogen molecule (See Supplementary Info Sec. V), where $r_w$ is the effective radius. Using this model, it is possible to estimate the number of adsorption sites $B$ as the ratio of the pore surface area to the cross sectional area of a hydrogen molecule ($\pi r_w^2$), which leads from (6) to the adsorption formula for a material given a unit cell (7).

Strictly speaking we should also consider edge effects when one hydrogen molecule occupies two or more different sites. However, edge effects are significant only for frameworks, in which the linker length is nearly the same as the diameter of the guest molecule. In this case, most known synthesized frameworks have linker lengths that are significantly larger than the hydrogen molecule diameter, thus the edge effects should be negligible for current synthesized materials. Finally, we obtain the equation for uptake, which can be applied to any type of framework with only one type of conventional unit cell.

$$m = \frac{\gamma\mu V}{N_A\pi r_w^2}\frac{S_p}{V_p}{}_{N/B}\left(\tilde{n}\right). \tag{7}$$

$S_p$ and $V_p$ in equation (7) are pore surface area and pore volume, respectively. The occupancy ratio $N(\tilde{n})/B$ is defined using the RMT (5) by assuming that $j_1$ and $j_\infty$ are linked with enthalpies of adsorption on the first layer $\Delta H_{\text{ads}}$ and higher layers $\Delta H_{\text{ads}}^*$ as follows (for more details see the Supplementary Info Sec. I).

$$j_1 = \exp(\Delta H_{\text{ads}}/RT), \ j_\infty = \exp(\Delta H_{\text{ads}}^*/RT). \tag{8}$$

The enthalpy of adsorption is crucial for the application of the RMT. If the enthalpy of adsorption is known, the theory is a significantly accurate model. When comparing RMT to experimental results on uptake we were using experimentally available data on enthalpy of adsorption $\Delta H_{\text{ads}}$ for MOFs. However, as reported by Mendoza-Cortes et al[5], the $\Delta H_{\text{ads}}$ is almost equal to binding energy of hydrogen to porous material site $\Delta H_{\text{bind}}$, and hence can also be calculated in a straightforward manner using Density Functional Theory (DFT).

We may generalize our derivation to $M$ types of conventional unit cells with volumes and surface areas $V_p^i$ and $S_p^i$, respectively. We also introduce factors $\gamma_1, \ldots, \gamma_M$ for the fraction of volume occupied by every type of the cell,

$$\gamma = \sum_{i=1}^{M} \gamma_i. \tag{9}$$

Applying equation (7) to every type of cell separately, we get the total uptake for the most general configuration given different types of cells,

$$m = \sum_{i=1}^{M} \frac{\gamma_i \mu V}{N_A \pi r_w^2} \frac{S_p^i}{V_p^i} \frac{N}{B} (\tilde{n}_i). \tag{10}$$

This final general equation can be used for the different porous materials reported in the literature that serves as hosts for many chemical compounds to form energy storage materials.

# 3    Results and discussions

**Comparison to experiments.**    To validate our results, we compared our predictions to experimental data for different reported MOFs at $T = 77$ K. We took published data from six reported frameworks, namely MOF-5, MOF-177, MOF-14, SNU-77H and MOF-200 for comparison over geometric properties and adsorption enthalpy (however, DFT calculations can be used to estimate these values as well[5]). We present the theoretical results and compare them to experiment for these frameworks in Table 1. The agreement between experimental and theoretical uptake is within 4 % at this temperature. RMT shows similar average absolute error (0.89 g/L) in excess uptake to Langmuir (0.90 g/L) and BET (0.77 g/L) theories when using the uptake formula (10) with $\tilde{n} = 1$ and $\tilde{n} = \infty$ respectively. In Figure 3, we also compared the experimental and theoretical full isotherm at $T = 77$ K for MOF-5 and MOF-177. We find some discrepancies in the pressure range of 5-30 bar. This can be explained by the slight decrease of adsorption enthalpy reported in Wong-Foy et al,[6] thus suggesting that an effective $\Delta H_{ads}$ should be used rather that the one at low coverage as reported in experiments (see Supplementary Info Sec. V). Otherwise, there is significant agreement between the RMT results and experiment.

We also compare our theory with Grand Canonical Monte Carlo (GCMC) simulations for Li-metalated frameworks. Han et al.[7] considered Li-metalated frameworks with high enthalpy of adsorption at room temperature. This enthalpy decreases significantly with increasing uptake. Taking the enthalpy of adsorption reported by Han et al.[7] as the value for $P = 1$ bar, we took the change in enthalpy with pressure into account and defined an effective enthalpy (see Supplementary Info Sec. V) for the various pressures. We made an assumption that introducing lithium does not significantly influence the geometric properties of the framework, as suggested in Han et al,[7] and therefore we used the same geometric parameters of the non-metalated frameworks. The theoretical results for the three different temperatures $T = 243$ K, $T = 273$ K and $T = 298$ K were compared with simulations (see Figure 4). The theoretical predictions are in good agreement with the GCMC results over the entire pressure range. The average absolute deviation for two compounds are reported in Table 2; this error does not exceed 2.2 %.

Table 1: Experimental and theoretical *excess* uptake for different MOFs given their topology. $\Delta H_{ads}$ is given as the enthalpy of adsorption, $a$ is the characteristic length of the cell in a framework, and $S_A$ is the surface area.

| MOF | $a$ Å | $V^{\text{void}}$ $\gamma$ | $\tilde{n}$ | $S_A$ m²/g | $V_p$ cm³/g | $\Delta H_{ads}$ kJ/mol | $P$ bar | Exp. g/L | Theor. g/L | Err. % |
|---|---|---|---|---|---|---|---|---|---|---|
| MOF-5 | 12[8] | 0.798 | 2 | 2900[8] | 1.04[8] | 4.8[9] | 50 | 31.0[6] | 31.22 | +0.7 |
| MOF-14 | 8, 14[10] | 0.67[10] | 2 | 2000[10] | 0.71[11] | 7.0[11] | 40 | 28.1[11] | 27.25 | -3.0 |
| SNU-77H | 8[12] | 0.69[12] | 2 | 3900[13] | 1.52[13] | 7.05[13] | 90 | 47.4[13] | 49.26 | +3.9 |
| MOF-200 | 18[8] | 0.90[14] | 3 | 6400[14] | 3.59[14] | 3.46[15] | 90 | 16.5[14] | 17.18 | +4.0 |
| MOF-177 | 8.5[6] | 0.83[8] | 1 | 4740[8] | 1.59[13] | 4.4[6] | 70 | 32.0[6] | 31.17 | -2.6 |

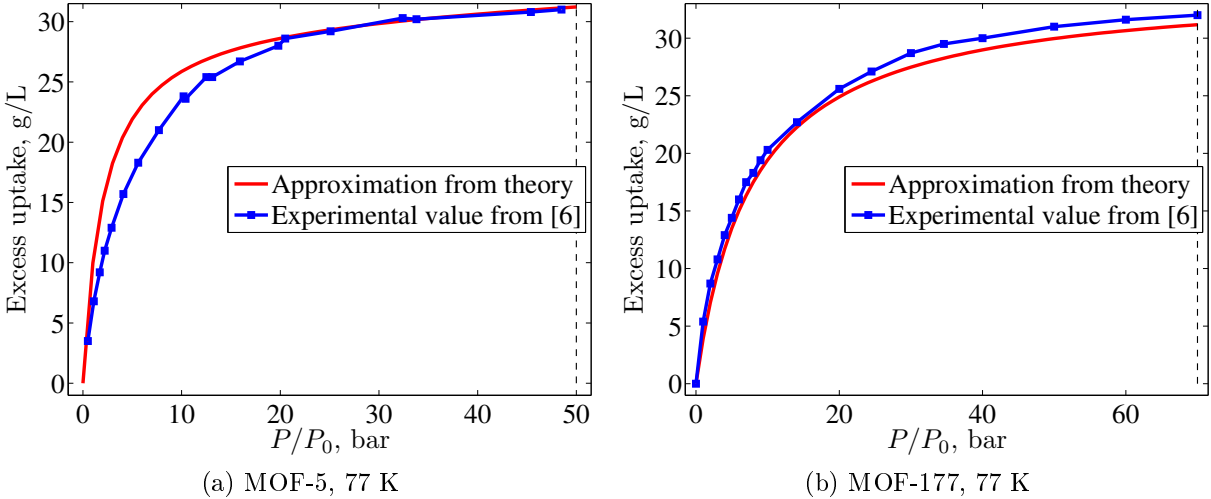

(a) MOF-5, 77 K    (b) MOF-177, 77 K

Figure 3: Comparison of theory and experiments for hydrogen isotherm in two porous materials at $T$=77 K.

Table 2: Average absolute deviation of excess uptake in g/L(%) for MOF-177-15Li and MOF-200-27Li

| Compound | $T = 243$ K | $T = 273$ K | $T = 298$ K |
|---|---|---|---|
| MOF-177-15Li | 1.80 (6.0%) | 2.12 (7.1%) | 1.67 (5.6%) |
| MOF-200-27Li | 1.02 (5.0%) | 1.21 (6.0%) | 0.40 (2.0%) |

**Optimal parameters of the framework for the highest gas delivery using the RMT.** The ultimate goal of making porous materials for storing hydrogen is not only to store as much molecular hydrogen as possible at some fixed pressure $P$ (say 100 bar), but also to deliver the maximum amount of hydrogen to the surroundings upon depressurization, which is known as the net delivery amount of the storage device. Analytically, this is defined as the difference in uptake between the atmospheric pressure, $P_0$ (usually 1 bar), and the maximum pressure of the tank, $P$ (say 100 bar). Since minimizing the amount of stored hydrogen lost on delivery is the goal, obtaining the maximum net delivery amount is the most important characteristic of a storage device. We propose doing this by obtaining an optimal enthalpy of adsorption $\Delta H_{ads}^{opt}$. Porous materials with low $\Delta H_{ads}$ do not store enough gas, even at the maximum pressure $P$. On the other hand, if $\Delta H_{ads}$ is too high the framework retains all the adsorbed gas at low pressure, and the gas cannot be used.[16] In both cases, the delivery amount is negligible or very low. Therefore, we first derive equations for optimal enthalpy of adsorption ($\Delta H_{ads}^{opt}$) using RMT. Then, we present an approach of determining the optimal geometric properties (i.e. effective linker length) for a given series of frameworks.

**Optimal Enthalpy of adsorption $\Delta H_{ads}^{opt}$.** First, we find optimal values of $\Delta H_{ads}^{opt}$ in a low-temperature region (for instance, for $T = 77$ K), where rotational degrees of freedom are not significantly activated. Nevertheless this theory is further applied for high temperatures (for instance, for $T = 298$ K) just by considering the full chemical potential (see Supplementary Info Sec. VI). One may use the full chemical potential for the low temperature region as well; however, in this region, the equilibrium ratio of para- and ortho- hydrogen is far from 3:1. This implies full quantum mechanical considerations of ortho- and para- hydrogen without simplification (see Suppl. Info Sec. VI). However, this does not change the result significantly.

We derive the final expressions for the optimal enthalpy for the restricted multilayer case in Suppl. Info Sec. ID and find it is given by eq. (11).

$$\Delta H_{ads}^{opt} = RT \ln \beta_1 = RT \ln \tilde{\beta}_1/\alpha, \text{ where} \tag{11}$$

$$\tilde{\beta}_1 = \frac{a(1 - m\tilde{\beta}_2) - b\sqrt{m}(1 - \tilde{\beta}_2)}{b\sqrt{m}(1 - \tilde{\beta}_2^n) - am(1 - m^n\tilde{\beta}_2^n)}, \tag{12}$$

$$a = \sqrt{1 - (n+1)\tilde{\beta}_2^n + n\tilde{\beta}_2^{n+1}}, \ b = \sqrt{1 - (n+1)m^n\tilde{\beta}_2^n + nm^{n+1}\tilde{\beta}_2^{n+1}}. \tag{13}$$

We define

$$\beta_1 = \exp\{\Delta H_{ads}/(RT)\}, \ \beta_2 = \exp\{\Delta H_{ads}^*/(RT)\}, \ \tilde{\beta}_1 = \alpha\beta_1, \ \tilde{\beta}_2 = \alpha\beta_2, \tag{14}$$

$$c = \beta_1/\beta_2, \ x = \alpha P/P_0\beta_2, \tag{15}$$

where $\alpha$ is a function of temperature characterized further in Supplementary Information (See Supplementary Information Section VI, VII for derivation).

For low temperatures, the chemical potential $\mu$ is defined by a translational part, so

$$\alpha = \frac{P_0}{kT}\left(\frac{\mu kT}{2\pi N_A \hbar^2}\right)^{-3/2}. \tag{16}$$

For high temperatures, when the rotational degrees of freedom are fully activated,

$$\alpha = \frac{P_0}{kT}\left(\frac{\mu kT}{2\pi N_A \hbar^2}\right)^{-3/2}\left(\frac{2T_r}{T}\right). \tag{17}$$

Particularly for the high-temperature region, (e.g. $T = 298$ K), and the associated value of $\alpha$, we get

$$\Delta H_{ads}^{opt} = RT \ln \beta_1 = RT \ln \left(\tilde{\beta}_1 T^{7/2} / \left(\alpha_0 T_0^{7/2}\right)\right), \tag{18}$$

where $\alpha_0 = 5.1 \cdot 10^{-6}$, $T_0 = 298$ K and

$$\tilde{\beta}_1 = \alpha_0 \left(T_0/T\right)^{7/2} \beta_1. \tag{19}$$

Thus for the RMT, for example at $T_0 = 77$ K, $P = 100$ bar, $n = 3$, we have $\Delta H_{ads}^{opt} = 3.81$ kJ/mol. On the other hand, at $T_0 = 298$ K, $P = 100$ bar, $n = 3$, we have $\Delta H_{ads}^{opt} = 24.5$ kJ/mol.

The optimal $\Delta\nu$ is defined as $\nu(P) - \nu(P_0)$ where $\tilde{\beta}_1$ is denoted optimal given by Eq. 12. The optimal $\Delta\nu$ for $n = 1$ or the Langmuir case can be obtained if $\tilde{\beta}_2 = 0$ and therefore $a = b = 1$ in the previous equations. Similarly, the optimal $\Delta\nu$ for $n = \infty$ or the BET case can be derived by excluding powers of $\tilde{\beta}_2$ higher than first power and setting $a = b = 1$. This is another example where the Restricted Multilayer Theory is the general case of the Langmuir and BET theories.

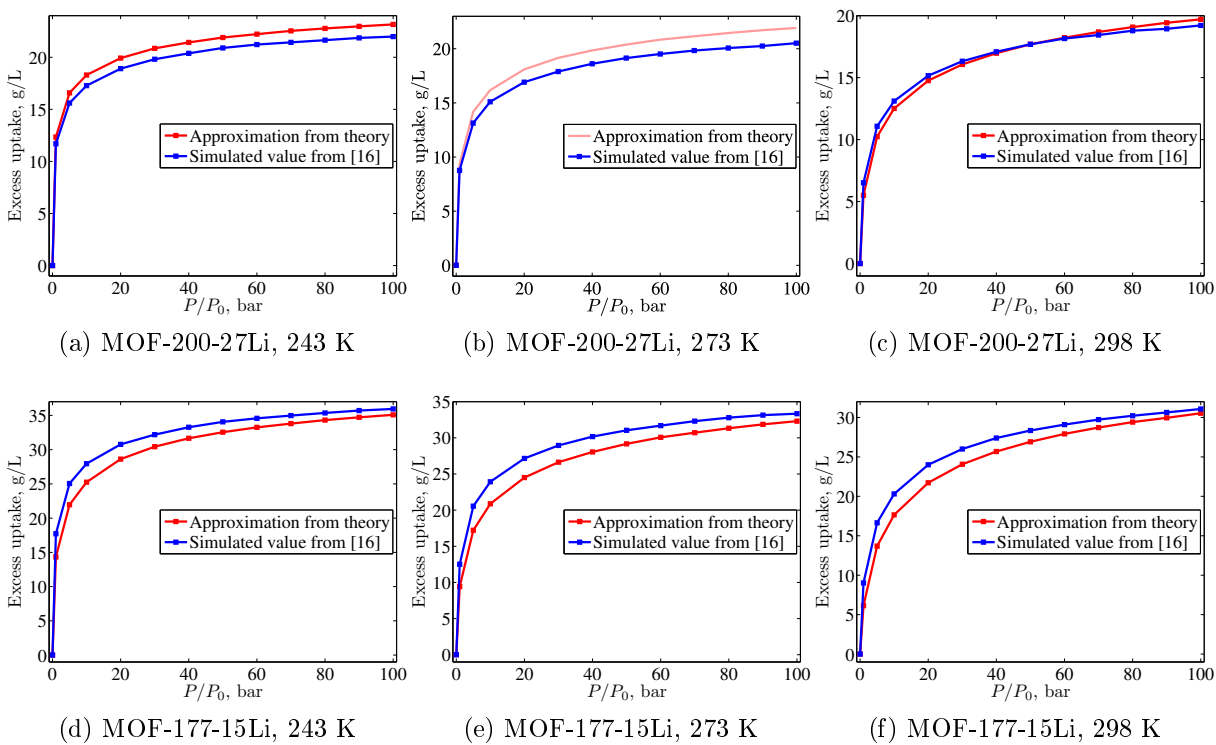

(a) MOF-200-27Li, 243 K     (b) MOF-200-27Li, 273 K     (c) MOF-200-27Li, 298 K

(d) MOF-177-15Li, 243 K     (e) MOF-177-15Li, 273 K     (f) MOF-177-15Li, 298 K

Figure 4: Theoretical RMT and Grand Canonical Monte Carlo simulated adsorption isotherms for (a-c) MOF-200-27Li and (d-f) MOF-177-15Li at 243, 273, and 298 K

Now, to illustrate the consequences of these equations we first plot optimal enthalpy at different temperatures and different number of layers ($n = 2, 3, 6$) in Figure 5 and show some of the values in Table 3.

This table shows a similar performance for different amounts of layers. Thus, using RMT, the figures are similar for any number of layers, therefore, we only show the plots for $n = 2$. Discrepancies are noticeable only for very low temperatures and for large differences in the

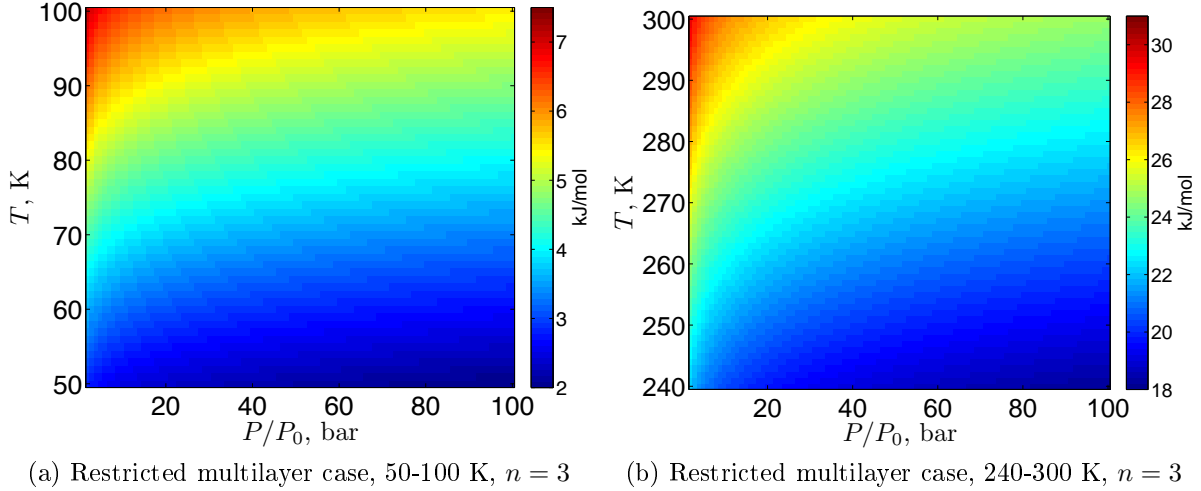

(a) Restricted multilayer case, 50-100 K, $n = 3$

(b) Restricted multilayer case, 240-300 K, $n = 3$

Figure 5: Temperature and pressure dependence of optimal enthalpy $\Delta H_{ads}^{opt}$ for low and high temperature region and $n$=3.

Table 3: Optimal enthalpy $\Delta H_{\mathrm{ads}}^{opt}$ for different temperature $T$ and number of layers $n$ for $P = 100$ bar

| Number of layers $n$ | $T = 50$ K | $T = 77$ K | $T = 298$ K |
|:---:|:---:|:---:|:---:|
| 1 | 2.02 | 3.80 | 24.48 |
| 2 | 1.99 | 3.80 | 24.48 |
| 3 | 1.99 | 3.81 | 24.48 |
| 6 | 1.99 | 3.81 | 24.48 |

number of layers. From eq. 11 we find that the optimal enthalpy increases with temperature.

Finally, we plot the results for the optimized value for $\Delta \nu = \Delta(N/B)$ which directly influences the shape of the isotherm (see Figure 6). These results show that the difference between all three cases becomes significant only at low temperature. However, one must be cautious for other gases, which have higher $\Delta H_{ads}^{opt}$ than hydrogen and this difference can be much more significant when applying the different theories. The safe way, of course, is to use RMT.

**Optimal linker length and void volume.** In the previous section, we present the optimal value of enthalpy which influences the shape of the isotherm. Here, we present an approach to define optimal linker length for a series of frameworks. The linker length does not influence the shape of the curve, but it does influence its asymptotic value (when $P \to \infty$).

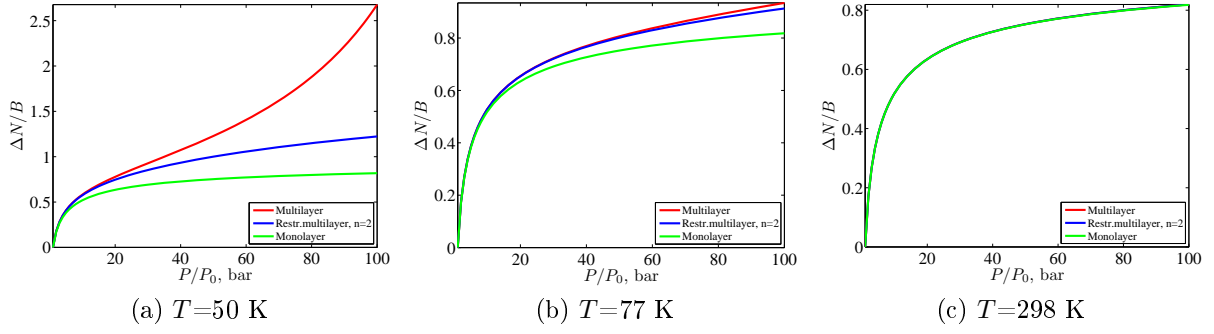

(a) $T$=50 K  (b) $T$=77 K  (c) $T$=298 K

Figure 6: The ratio $\Delta N/B$ for different temperatures

This is because the pore volume $V_p$, surface area $S_p$ and fraction of void volume $\gamma$ depend on the linker length.

We derive an expression to determine optimal linker length and void volume in Suppl. Info Sec. V and get

$$\gamma = 1 - \frac{V_{net}}{V} = 1 - \frac{\kappa}{a^{2-\xi}}, \tag{20}$$

where $\kappa$ is some constant depending on the framework and $\xi$ indicates how the volume grows with increasing $a$. We also get $S_p/V_p = c/a$, where $c$ is constant for every framework not changing when enlarging the linkers.

Thus,

$$f(a) = \gamma \frac{S_p}{V_p} = \left(1 - \frac{\kappa}{a^{2-\xi}}\right) \frac{c}{a} \to \text{Maximize } f(a) \tag{21}$$

This gives

$$a^{2-\xi} = (3-\xi)\kappa \tag{22}$$

and therefore

$$a = ((3-\xi)\kappa)^{\frac{1}{2-\xi}}. \tag{23}$$

The factor $\kappa$ can be easily defined from experimental data if $\gamma$ is known for at least one length of the linker $a$ or can be measured directly for any framework just by using geometry. Therefore, we obtain the universal recipe of best linker length under these assumptions.

To show an example, given $\xi = 0$, for any framework the best value is

$$\gamma = 1 - \frac{\kappa}{(\sqrt{3\kappa})^2} = 2/3. \tag{24}$$

Now we apply equation (23) to the topologically equivalent IR-MOFs, IRMOF-1 to IRMOF-16. This will determine the optimal length of the linker for such frameworks and therefore the optimal void volume. First we determine the model parameters $\kappa$ and $\xi$. Rearranging Eq. 20 we obtain $\log(1 - \gamma) = \log \kappa - (2 - \xi) \log a$. Now, using data over the fraction of void volume $\gamma$ and the effective linker length $V_p/S_p$ for the considered IRMOFs, we obtain $\xi$ and $\kappa$ using a linear regression (Figure 7a). The regression has a high correlation coefficient, $R^2 = 0.9524$ showing the significance of the dependence of the void volume and the effective linker length. Thus for this family of MOFs, we obtain $\log \kappa = -0.015$, $2 - \xi = 0.932$ and the estimations for the coefficients are $\kappa = 0.966$, $\xi = 1.068$.

Thus we get the optimal void volume $\gamma = 0.4824$. and the optimal effective linker length $(V_p/S_p) = 1.95$ Å.

Note that the effective linker length is not the same as the linker length alone. For instance, for a cubic cell, the effective linker length is $a^3/(6a^2)$ which is 6 times smaller than the linker length $a$. The closest one to the reported figures is IRMOF-5 with $\gamma = 0.5$ and $a = 2.05$ Å, which correspond to 98.5% of the predicted optimal uptake for this framework.

The optimal quantity for effective linker length is close to the value for IRMOF-5 and therefore further improvements for the IRMOFs should not consider the linker length but instead the enthalpy of adsorption (24.5 kJ/mol, for $n = 1$ and $T = 298$ K, given that not more than one layer of hydrogen can be inserted in such pores).

# 4    Conclusions

We presented a theory for restricted multilayer adsorption of gases in crystalline porous materials called the Restricted Multilayer Theory (RMT). We find this model for adsorption to be realistic since having one or infinite layers (as Langmuir and BET assumes) is unphysical and restricted multilayer theory is more applicable for solving problems such as storing molecular hydrogen in porous materials.

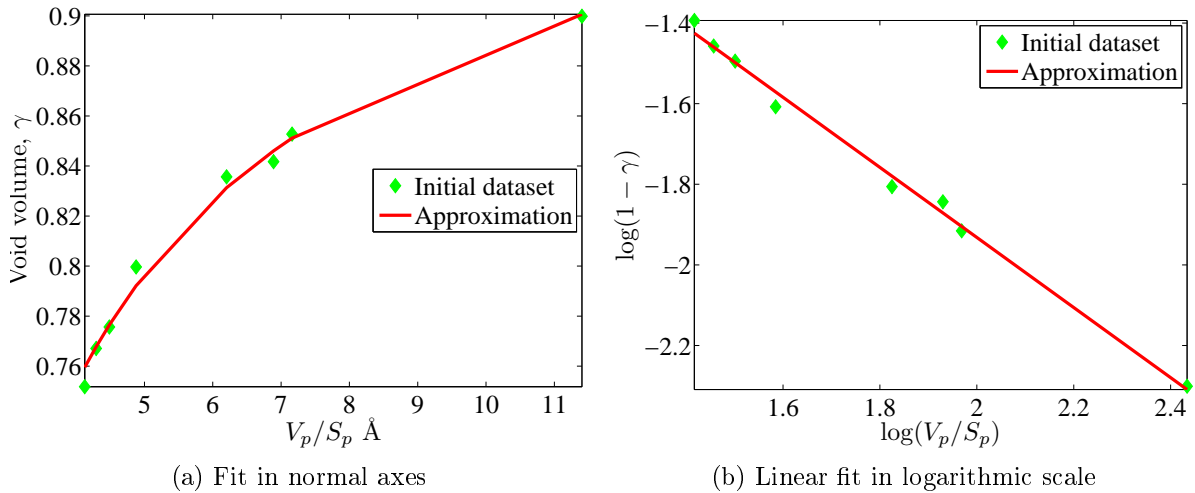

(a) Fit in normal axes      (b) Linear fit in logarithmic scale

Figure 7: Void volume dependence on linker length. Notice the logarithmic axes in (b) and its direct correlation

RMT fixes several inconsistencies of the Langmuir and BET theory. For example, both BET and Langmuir guarantee at least one adsorption layer which is not physical for gases with molecules bigger than the pore size. The BET theory has an intrinsic inconsistency as if there is such pressure $P(T)$, so that when $P \to P(T)$ the uptake becomes infinite, which is unphysical, e.g. for 77K this pressure is 955 bar.

The occupancy ratio for the RMT is characterized by equation 5 which allows for a restricted number of layers $n$. We applied this theory to estimate molecular hydrogen adsorption in Metal-Organic Frameworks (MOFs) and our approximations predict uptakes that agree with experimental values at 77 K and 298 K. We found that the results for this theory can differ from the Langmuir and BET theories at low temperatures and that the RMT is a better predictive tool at these temperatures.

For compounds that are isoreticular to MOF-5 (topology **pcu**), we found that the optimal effective linker length is 1.95 Å and the enthalpy of adsorption is 3.75 kJ/mol at 77 K and 24.5 kJ/mol at 298 K, respectively, which will give a maximum expected delivery amount of 81.8% or 34.65 g/L. The RMT can be used to estimate the optimal parameters before making a porous material and thus will be very useful tool in the future design of materials for fuels and other energy storage.

**Supporting Information.** Full derivations for Restricted Multilayer Isotherm, gas uptake from RMT, optimal enthalpy, optimal linker length and void volume, are included.

**Acknowledgements** J.L.M-C. were supported by Florida State University (FSU) and its Energy and Materials Initiative. This paper was submitted on the birthday anniversary of Irving Langmuir.

# References

(1) Langmuir, I. The Constitution and Fundamental Properties of Solids and Liquids, Part I. Solids. *Journal of the American Chemical Society* **1916**, *38*, 2221–2295.

(2) Brunauer, S.; Emmett, P. H.; Teller, E. Adsorption of Gases in Multimolecular Layers. *Journal of the American Chemical Society* **1938**, *60*, 309–319.

(3) Hill, T. L. Statistical Mechanics of Multimolecular Adsorption. I. *The Journal of Chemical Physics* **1946**, *14*, 263–267.

(4) Keii, T. A Statistical Derivation of the B.E.T. Equation. *The Journal of Chemical Physics* **1954**, *22*, 1617–1618.

(5) Pramudya, Y.; Mendoza-Cortes, J. L. Design Principles for High H2 Storage Using Chelation of Abundant Transition Metals in Covalent Organic Frameworks for 0–700 bar at 298 K. *Journal of the American Chemical Society* **2016**, *138*, 15204–15213.

(6) Wong-Foy, A. G.; Matzger, A. J.; Yaghi, O. M. Exceptional H2 Saturation Uptake

in Microporous Metal-Organic Frameworks. *Journal of the American Chemical Society* **2006**, *128*, 3494–3495.

(7) Han, S. S.; Jung, D. H.; Choi, S.-H.; Heo, J. Lithium-Functionalized Metal-Organic Frameworks that Show >10 wt % H2 Uptake at Ambient Temperature. *ChemPhysChem* **2013**, *14*, 2698–2703.

(8) Eddaoudi, M.; Moler, D. B.; Li, H.; Chen, B.; Reineke, T. M.; O'Keeffe, M.; Yaghi, O. M. Modular chemistry: secondary building units as a basis for the design of highly porous and robust metal- organic carboxylate frameworks. *Accounts of Chemical Research* **2001**, *34*, 319–330.

(9) Rowsell, J. L. C.; Yaghi, O. M. Effects of Functionalization, Catenation, and Variation of the Metal Oxide and Organic Linking Units on the Low-Pressure Hydrogen Adsorption Properties of Metal-Organic Frameworks. *Journal of the American Chemical Society* **2006**, *128*, 1304–1315.

(10) Chen, B.; Eddaoudi, M.; Hyde, S.; O'Keeffe, M.; Yaghi, O. Interwoven Metal-Organic Framework on a Periodic Minimal Surface with Extra-Large Pores. *Science* **2001**, *291*, 1021–1023.

(11) Han, S. S.; Mendoza-Cortés, J. L.; Goddard III, W. A. Recent advances on simulation and theory of hydrogen storage in metal-organic frameworks and covalent organic frameworks. *Chemical Society Reviews* **2009**, *38*, 1460.

(12) Park, H. J.; Lim, D.-W.; Yang, W. S.; Oh, T.-R.; Suh, M. P. A Highly Porous Metal-Organic Framework: Structural Transformations of a Guest-Free MOF Depending on Activation Method and Temperature. *Chemistry - A European Journal* **2011**, *17*, 7251–7260.

(13) Suh, M. P.; Park, H. J.; Prasad, T. K.; Lim, D.-W. Hydrogen Storage in Metal-Organic Frameworks. *Chemical Reviews* **2012**, *112*, 782–835.

(14) Furukawa, H.; Ko, N.; Go, Y. B.; Aratani, N.; Choi, S. B.; Choi, E.; Yazaydin, A. Ö.; Snurr, R. Q.; O'Keeffe, M.; Kim, J. Ultrahigh Porosity in Metal-Organic Frameworks. *Science* **2010**, *329*, 424–428.

(15) Mendoza-Cortés, J. L.; Han, S. S.; Goddard, W. A. High H2 Uptake in Li-, Na-, K-Metalated Covalent Organic Frameworks and Metal Organic Frameworks at 298 K. *The Journal of Physical Chemistry A* **2012**, *116*, 1621–1631.

(16) Pramudya, Y.; Mendoza-Cortes, J. L. Design Principles for High H2 Storage Using Chelation of Abundant Transition Metals in Covalent Organic Frameworks for 0-700 bar at 298 K. *Journal of the American Chemical Society* **2016**, *138*, 15204–15213.

# Graphical TOC Entry

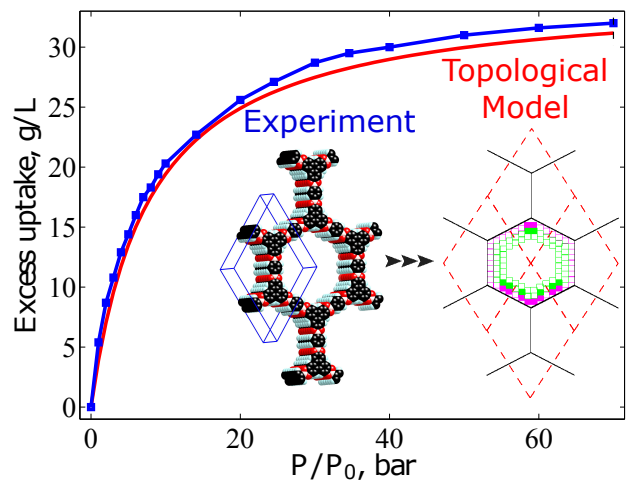

# Supporting Information

# General Theory of Absorption in Porous Materials: The Restricted Multilayer Theory

Alexander A. Aduenko[1], Andy Murray[2], and Jose L. Mendoza-Cortés[*,2,3],

[1] Department of Control and Applied Mathematics, Moscow Institute of Physics and Technology, Dolgoprudny, 141700, Russia
[2] Department of Scientific Computing, Materials Science and Engineering Program, High Performance Materials Institute, Condensed Matter Theory - National High Magnetic Field Laboratory, Tallahassee FL, 32310, USA
[3] Department of Chemical & Biomedical Engineering, FAMU-FSU Joint College of Engineering, Tallahassee FL, 32310, USA
[4] Department of Physics, College of Arts and Science, Florida State University, Tallahassee FL, 32310, USA.

E-mail: mendoza@eng.famu.fsu.edu

## I. ADSORPTION THEORY

In this subsection we firstly derive the Langmuir or monolayer theory of adsorption. Following this, we consider Brunauer-Emmett-Teller (BET) or multilayer theory. Finally, we derive a more general Restricted Multilayer Theory (RMT). Langmuir and BET theories are RMT's special cases, when the number of possible adsorption layers equal 1 and ∞ respectively. Finally, we will see that we are able to overcome the unphysical restrictions imposed by Langmuir and BET theories, i.e. the requirement of a monolayer for the Langmuir theory and infinite layers for the BET case.

We make the following definition and assumptions based on the postulates above. In the system, there are $B$ equivalent sites, there are $N$ molecules distributed in $B$ sites, there are no interactions between molecules and the lattice may be one-, two-,three-periodic.

### A. Derivations of the different absorption theories

***Monolayer theory.*** Using the the Gibbs ensemble (also known as Grand Canonical distribution), we have

$$\Xi(\mu, \tau) = \sum_{N=0}^{\infty} \sum_{S(N)} \exp[(N\mu - \epsilon_{S(N)})/\tau]$$
$$= \sum_{[S(N),N]} \exp[(N\mu - \epsilon_{S(N)})/\tau],$$

where,

- $[S(N), N]$ is for all states of the system for all number of particles

- $\epsilon_{S(N)}$ defines that each $\epsilon_S$ depends on the number of particles. $\epsilon_{S(N)}$ is the energy of the state $S(N)$ of the exact $N$-particle Hamiltonian.

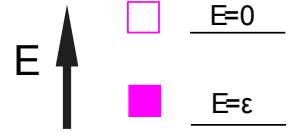

FIG. 1: Assumption from Langmuir theory

Now we can calculate the thermal average number of particles by

$$\overline{N} = \frac{\sum_{S,N} N \exp[(N\mu - \epsilon_{S(N)})/\tau]}{\Xi}.$$

Combining that with

$$\frac{\partial \Xi}{\partial \mu} = \frac{1}{\tau} \sum_{S,N} N \exp[(N\mu - \epsilon_{S(N)})/\tau]$$

we obtain:

$$\overline{N} = \frac{\tau}{\Xi} \frac{\partial \Xi}{\partial \mu} = \tau \frac{\partial \ln \Xi}{\partial \mu}.$$

Here we considered only one adsorption site. Let $B$ be the number of adsorption sites. From the definition of $\lambda = \exp(\mu/\tau)$, we have

$$\frac{\overline{N}}{B} = \lambda \frac{\partial \ln \Xi}{\partial \lambda} \qquad (1)$$

Now we take into account the assumption from the Langmuir theory as two energy states for each particle, as shown in the figure 1. We consider adsorption as favorable process. Therefore, the lower energy corresponds to the adsorbed state, and thus $\epsilon < 0$. Given this, we can derive the partition function as

$$\Xi = 1 + \lambda \exp\left(\frac{-\epsilon}{\tau}\right)$$

Substituting this result into eq. (1), we have

$$\frac{\overline{N}}{B} = \frac{\lambda \exp(-\epsilon/\tau)}{1 + \lambda \exp(-\epsilon/\tau)}$$

$$\frac{\overline{N}}{B} = \frac{1}{\lambda^{-1}\exp(\epsilon/\tau) + 1} \tag{2}$$

**Multilayer theory.** It is postulated that a gas molecule can be absorbed on each site $B$, with partition function $j_1$, which is denoted as the "first layer". This first molecule can then be used as a site for another "second layer" molecule, and so on. The partition function for the second and subsequent molecules is given as $j_\infty$. If given $N$ total number of molecules, of which $N_1$ are in the first layer, then $N^* = N - N_1$ are in higher layers (more details can be found elsewhere [1], [2]).

Using the grand canonical partition function, we have

$$\Xi(B, \mu, T) = \sum_{N \geq 0} \exp[(N\mu)/k\tau] Q(N, B, T)$$

$$= 1 + \sum_{N_1=0}^{B} \frac{B!(j_1 \exp(\mu/(kT)))^{N_1}}{N_1!(B-N_1)!(N_1-1)!}$$

$$\times \sum_{N^*=0}^{\infty} \frac{(N_1 + N^* - 1)!(j_\infty \exp(\mu/(kT)))^{N^*}}{N^*!}$$

Furthermore,

$$\Xi(B, \mu, T) = \sum_{N_1=0}^{B} \frac{B!(y)^{N_1}}{N_1!(B-N_1)!} = (1+y)^B \tag{3}$$

where we have defined,

$$y = \frac{j_1 \exp(\mu/kT)}{1 - j_\infty \exp(\mu/kT)}$$

Again using

$$\overline{N} = kT\left(\frac{\partial \log \Xi}{\partial \mu}\right)_{T,B} \tag{4}$$

and applying it to equation 3, we have

$$\frac{\overline{N}}{B} = \frac{cx}{(1 - x + cx)(1 - x)} \tag{5}$$

where $c = \frac{j_1}{j_\infty}$ and $x = j_\infty \exp(\mu/kT)$. This is also known as the BET adsorption isotherm equation [3].

**Restricted Multilayer theory (RMT)** A more realistic approach is that the adsorption is restricted to $n$ layers. We assume that there are only three types of interactions in this system, which are

1. Those between the surface ($n = 0$) and the first layer molecules ($n = 1$), with partition function $j_1$
2. Those between first layer molecules ($n = 1$) and second layer molecules ($n = 2$), with partition function $j_2$, and
3. Those between $n^{\text{th}}$ ($n > 1$) layer molecules and $n^{\text{th}} + 1$ layer molecules, with partition function $j_\infty$.

This creates a columnar structure of bound molecules. Additionally, we assume that there are no interactions between molecules of neighboring columns and the partition function $j_2$ is equal to $j_\infty$. Given the total number of molecules, $N$, then $N_1$ of them are in the first layer and $N_n = N - N_1$ are in higher layers. From these set of assumptions, we can derive the adsorption behavior for a restricted number of multilayer.

Using the grand canonical partition function, we have

$$\Xi(B, \mu, T) = \sum_{N \geq 0} \exp[(N\mu)/kT] Q(N, B, T)$$

$$= \sum_{N_1=0}^{B} \sum_{N_2=0}^{N_1} \sum_{N_3=0}^{N_2} \cdots \sum_{N_n=0}^{N_{n-1}} \frac{B!}{(B-N_1)!N_1!}$$

$$\times \frac{N_1!}{(N_1 - N_2)!N_2!} \cdots$$

$$\times \frac{N_{n-1}!(cx)^{N_1} x^{N_2 + \ldots + N_n}}{(N_{n-1} - N_n)!N_n!}.$$

In the preceding equations, $\mu$ is chemical potential, $T$ is temperature and $k$ is the Boltzmann constant.

Summing in turn over $N_n$, $N_{n-1}$, ..., $N_1$ we find

$$\Xi = [1 + cx(1 + x + x^2 + \ldots + x^{n-1})]^B$$

$$= \left[1 + cx\left(\frac{1-x^n}{1-x}\right)\right]^B$$

where we have defined,

$$c = j_1/j_\infty, \ \ x = j_\infty \exp(\mu/kT).$$

Using this equation in conjunction with the identity,

$$\overline{N} = kT\left(\frac{\partial \log \Xi}{\partial \mu}\right)_{T,B},$$

we find the equation for the *adsorption of a restricted number of multilayers*:

$$\frac{\overline{N}}{B} = \frac{cx[1 - (n+1)x^n + nx^{n+1}]}{(1-x)(1 - x + cx - cx^{n+1})}. \tag{6}$$

Source Code for the calculation of the Occupancy Ratio is found in Section IX Code 1.

This equation is more general, since assumptions on the number of layers $n = 1$ or $n = \infty$ give the Langmuir and BET equations, respectively. The substitution,

$$\frac{\overline{N}}{B} \xrightarrow{n=\infty} \frac{cx}{(1-x+cx)(1-x)}$$

gives the BET equation (Eq. 5), and

$$\frac{\overline{N}}{B} \xrightarrow{n=1} \frac{j_1 \exp(\mu/kT)}{1 + j_1 \exp(\mu/kT)}$$

gives the Langmuir equation (Eq. 2).

### B. Applications of these formulas to $H_2$ in porous materials

In this subsection we assume ideal behavior for molecular hydrogen gas and consider no interaction between its molecules. At this point, we neglect rotations, however consideration of rotations will be reported in the following sections. Changes in chemical potential, $\mu$, arising from rotations are fully derived later. Nevertheless, we should note that for $T = 77$ K rotations are not very significant, as the rotational activation temperature is 85.4 K for molecular hydrogen. Moreover, in order to consider rotations at such a temperature, one must separately consider orthohydrogen and parahydrogen because, at this temperature, their rotational partition functions differ significantly. Thus in this section we consider only translational part of entropy and free energy, which is exact for monoatomic gases, and is quite a good approximation for the hydrogen molecule at such a low temperature. In order to obtain consistent results for room temperature, we consider rotations later in this text. There is no issue with orthohydrogen and parahydrogen for room temperature as partition functions for both of them differ only slightly.

In this subsection we begin by applying our assumptions to the Langmuir or monolayer theory of adsorption. Following that, we consider the BET or multilayer theory. Finally, we employ the more general Restricted Multilayer Theory (RMT). Langmuir and BET theories are special cases of RMT, when the number of adsorption layers equals 1 and $\infty$, respectively.

According to prior assumptions, we get

$$\lambda = \frac{n}{n_q} = \frac{p}{\tau n_q} = \frac{p}{kT n_q}$$

and

$$n_q = \left(\frac{M\tau}{2\pi\hbar^2}\right)^{3/2} = \left(\frac{MkT}{2\pi\hbar^2}\right)^{3/2}$$

***Monolayer theory in molecular hydrogen.*** If we apply these assumptions to equation 2 we get the Langmuir equation for $H_2$

$$\frac{\overline{N}}{B} = \frac{1}{(n_q\tau/p)\exp(\epsilon/\tau) + 1}$$

$$\frac{\overline{N}}{B} = \frac{p}{(n_q\tau)\exp(\epsilon/\tau) + p}$$

$$= \frac{p/p_0}{(n_q\tau)\exp(\epsilon/\tau)/p_0 + p/p_0} \quad (7)$$

where $p_0 = 1$ bar.

***Multilayer theory in molecular hydrogen.*** In order to obtain $j_1$ and $j_\infty$, the procedure illustrated in Figure 2 was used. We consider adsorption as a favorable process and therefore the energies $\epsilon_1$ and $\epsilon_\infty$ are negative relative to the energy of a free molecule. Figure 2 illustrates the common situation when adsorption on higher levels is less favorable than adsorption on the first one and therefore $\epsilon_\infty > \epsilon_1$.

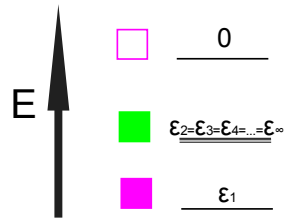

FIG. 2: Assumption for the multilayer theory

Thus for partition functions $j_1$ and $j_\infty$, for molecules adsorbed on first and higher levels, respectively, we get

$$j_1 = e^{-\epsilon_1/kT} = e^{\Delta H_{ads}/RT}$$

$$j_2 = j_3 = j_4 = ... = j_\infty = e^{-\epsilon_\infty/kT} = e^{\Delta H^*_{ads}/RT}$$

$$\Delta H_{ads}/RT = \text{Surface} \cdots H_2$$

$$\Delta H^*_{ads}/RT = \quad H_2 \cdots H_2$$

Figure 3 illustrates the adsorption process. Filled squares represent adsorbed molecules, and empty squares represent free sites. First layer molecules (colored in magenta) are bonding sites for higher level molecules (colored in green).

If we set $\epsilon_2 = 0$, then we will nearly have the monolayer case. There will be some effect due to the existence of higher levels, but it will not be significant in the studied range of pressure (0 to 100 bar). Consider

$$\alpha = \frac{P_0}{kT}\left(\frac{MkT}{2\pi\hbar^2}\right)^{-3/2},$$

where $P_0 = 1$ bar. For $T = 77$ K, $\alpha \approx 2.6 \cdot 10^{-4}$ which is a relatively small value. In the multilayer case for $H_2$

$$\frac{\langle N \rangle}{B} = \frac{\alpha \frac{P}{P_0} e^{\Delta H_{ads}/RT}}{1 + \left(e^{\Delta H_{ads}/RT} - e^{\Delta H^*_{ads}/RT}\right)\alpha \frac{P}{P_0}} \times$$
$$\frac{1}{1 - \alpha \frac{P}{P_0} e^{\Delta H^*_{ads}/RT}}, \quad (8)$$

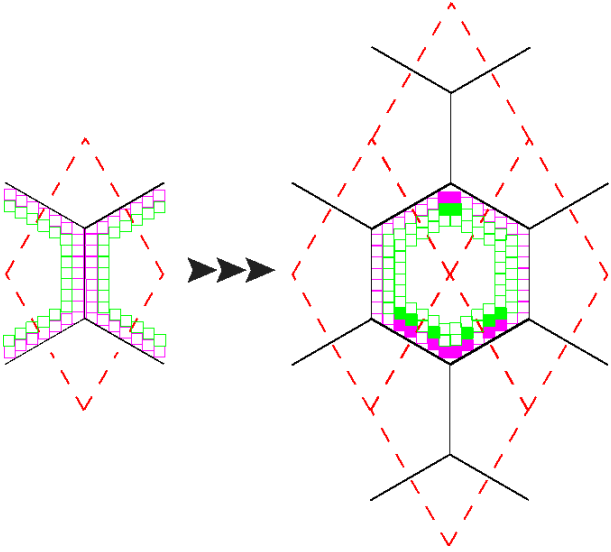

FIG. 3: Assumption for the multilayer theory

where $R = k N_A = 8.314 \, \text{J}/(\text{mol} \times \text{K})$. For hydrogen,

$$x = \alpha \frac{P}{P_0} \exp\left(\frac{\epsilon_2}{k_B T}\right),$$

and

$$c = \exp\left(\frac{\epsilon_1 - \epsilon_2}{k_B T}\right).$$

If we set $\Delta H_{ads}^* = 0$, then for the considered pressures

$$1 \geq 1 - \alpha \frac{P}{P_0} e^{\Delta H_{ads}^*/RT} \geq 1 - 2.6 \cdot 10^{-4} \times 100 \approx 1.$$

If $e^{\Delta H_{ads}^*/RT} \gg 1$, then

$$\frac{\langle N \rangle}{B} \approx \frac{1}{1 + \dfrac{1}{\alpha}\dfrac{P_0}{P} e^{-\Delta H_{ads}/RT}},$$

which is exactly the same as equation (7). When increasing $\Delta H_{ads}^*$, the effect from higher layers becomes more noticeable (compare Figure 4a and Figure 4b). Due to the limitation arising from $1 - x > 0$ in eq. (5), we get

$$\frac{\Delta H_{ads}^*}{RT} < -\log\left[\alpha \frac{P}{P_0}\right].$$

$\Delta H_{ads}^*$ is treated as equal to the heat of hydrogen liquefaction and therefore equals 0.904 kJ/mol.

The isotherm for the hydrogen molecule in the multilayer case for calculated $\Delta H_{ads}^*$ at $T = 77$ K is plotted (see Figure 4c). However, if $\Delta H_{ads}^*$ changes, the shape of the isotherm changes. For example, $\Delta H_{ads}^* = 0$ gives a similar plot to Figure 4c while $\Delta H_{ads}^* = 2 \, \text{kJ/mol}$ is very different (see Figure 4b vs Figure 4c).

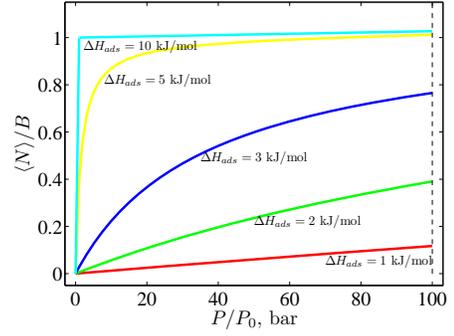

(a) $\Delta H_{ads}^* = 0 \, \text{kJ/mol}$

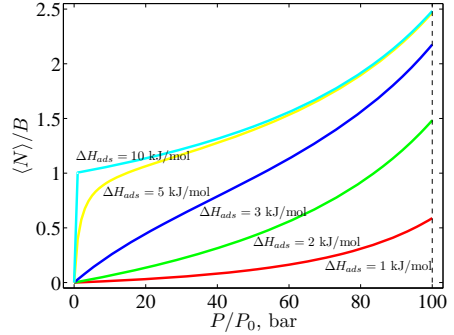

(b) $\Delta H_{ads}^* = 2 \, \text{kJ/mol}$

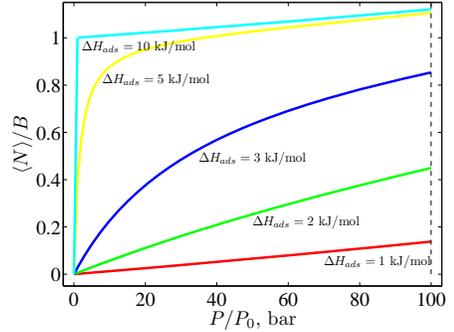

(c) Real case, $\Delta H_{ads}^* = 0.904 \, \text{kJ/mol}$

FIG. 4: Isotherms for $H_2$ in the multilayer case for $T = 77$ K

**Restricted Multilayer theory with molecular hydrogen.** Now we analyze molecular hydrogen through an application of RMT with $n$ layers. For $\epsilon_2 = 0$ (or $\Delta H_{ads}^* = 0$), we obtain that $x = \alpha P/P_0 \leq 2.6 \times 10^{-2}$. Moreover if $\Delta H_{ads} \leq 3 \, \text{kJ/mol}$, then $cx = \exp\left(\Delta H_{ads}/(RT)\right) x \leq 2.83$ for $T$=77 K. That is why, in this case, $cx^n$ has a small value for $n > 1$. But for $n = 1$ we have the monolayer case, which has already been considered. Furthermore, if we let $n \geq 2$, as $x \leq 2.6 \cdot 10^{-2}$ and $cx \leq 2.83$, we exclude from equation (6) all powers of $x$ higher than the first power and obtain

$$\frac{\langle N \rangle}{B} = \frac{cx}{1 + cx} \times \left(1 + \frac{2x}{1 + (c - 2)x}\right).$$

This relation shows that the difference between monolayer theory and restricted multilayer is not very significant if $\Delta H_{ads} \leq 3\,\text{kJ/mol}$, because the multiplying factor is the sum of 1 and a function raised to the first power of $x$. Now we consider isotherms of molecular hydrogen for different numbers of layers $n$, and different values of $\Delta H_{ads}$, depending on the relative pressure $P/P_0$.

For small values of $\Delta H^*_{ads}$, there is no significant dependence of $\langle N \rangle / B$ on the number of layers, whereas for high $\Delta H^*_{ads}$, this dependency is noticeable.

## C. Derivation for General Uptake Formula.

We first derive a general equation for the uptake, $m$, defined as the total mass of gas that can be incorporated per unit volume of a porous framework (in units of g/L). This equation applies for any framework based on the geometric constraints of the framework. Given that the molar mass of hydrogen $M = 2.018$ g/mol, the uptake per $V = 1\,L = 10^{27}\text{Å}^3$ equals

$$m = M \frac{N}{N_A},$$

where $N$ is the number of adsorbed molecules per unit of material volume, $N_A$ is the Avogadro constant. Let $B$ be the number of first layer sites per pore, then

$$m = M \frac{N}{N_A} = M \frac{BN/B(\tilde{n})n_p}{N_A}, \qquad (9)$$

where $n_p$ is the number of pores per unit of volume $V$, $N/B(\tilde{n})$ is the occupancy ratio defined by RMT theory with effective number of layers $\tilde{n}$ (6). Number of pores per unit of material volume $V$ is given by

$$n_p = \frac{\gamma V}{V_p}, \qquad (10)$$

where $\gamma$ is the fraction of void volume of the considered porous material, while $V_p$ is pore volume. Let $r_w$ be effective radius of adsorbed gas molecule meaning that one molecule occupies $\pi r_w^2$ of pore surface (see Section V for derivation of $r_w$ for molecular hydrogen), than neglecting edge effects it is possible to estimate the number of adsorption sites $B$ as the ratio of the pore surface area to the cross sectional area of a hydrogen molecule ($\pi r_w^2$)

$$B = \frac{S_p}{\pi r_w^2}, \qquad (11)$$

where $S_p$ is the pore surface area. Using (11) and (10) in (9) we get the following formula for uptake $m$

$$m = \frac{\gamma \mu V}{N_A \pi r_w^2} \frac{S_p}{V_p} N/B\,(\tilde{n}). \qquad (12)$$

We may generalize our derivation to $M$ types of conventional unit cells with volumes and surface areas $V_p^i$ and $S_p^i$, respectively. We also introduce factors $\gamma_1, \ldots, \gamma_M$ for the fraction of volume occupied by every type of the cell,

$$\gamma = \sum_{i=1}^{M} \gamma_i. \qquad (13)$$

Applying equation (12) to every type of cell separately, we get the total uptake for the most general configuration given different types of cells,

$$m = \sum_{i=1}^{M} \frac{\gamma_i \mu V}{N_A \pi r_w^2} \frac{S_p^i}{V_p^i} \frac{N}{B}(\tilde{n}_i). \qquad (14)$$

This final general equation can be used for the different porous materials reported in literature that serves as hosts for many chemical compounds to form energy storage materials.

## D. Derivation for Optimal Framework Parameters for the maximized gas delivery amount such as molecular hydrogen.

*Optimal enthalpy of absorption.* We find optimal values of $\Delta H_{ads}$ in a low-temperature region (for instance, for $T = 77$ K). We choose to investigate where rotational degrees of freedom are not significantly activated. We also pursue high temperatures (for instance, for $T = 298$ K) just by considering the full chemical potential later.

According to our theory, geometric properties of a framework only define some factors to the uptake. The shape of the isotherm is defined by the factor $\nu = N/B$ from the RMT. Thus to get the highest delivery amount for fixed number of layers (fixed geometry), one must derive the enthalpy $\Delta H_{ads}$ that leads to the highest possible difference in the factor $\nu$ between pressure $P_0$ and $P$.

$$\Delta \nu = [\nu(P) - \nu(P_0)] \rightarrow \text{Maximize } \Delta \nu.$$

The function $\nu(P)$ is smooth, thus the necessary condition to obtain the maximum is

$$\frac{\partial \Delta \nu}{\partial \Delta H_{ads}} \left( \Delta H^{opt}_{ads} \right) = 0. \qquad (15)$$

For convenience, we introduce the multiplier $m$ defined as $P = m\,P_0$

We also define

$$\beta_1 = \exp\{\Delta H_{ads}/(RT)\}, \ \beta_2 = \exp\{\Delta H^*_{ads}/(RT)\},$$

$$\tilde{\beta}_1 = \alpha \beta_1, \ \tilde{\beta}_2 = \alpha \beta_2,$$

$$c = \beta_1/\beta_2, \ x = \alpha P/P_0\beta_2.$$

Here, $H_{ads}^*$ is the interaction of first layer with the second layer of adsorbed molecules.

$$\alpha = \frac{P_0}{P} \exp(\mu/kT)$$

is a function of temperature. For the low temperature region chemical potential is defined by a translational part ($\mu \approx \mu_{tr}$) and

$$\alpha = \frac{P_0}{kT} \left( \frac{\mu kT}{2\pi N_A \hbar^2} \right)^{-3/2}.$$

For the high temperature region, when rotational degrees of freedom are fully activated,

$$\alpha = \frac{P_0}{kT} \left( \frac{\mu kT}{2\pi N_A \hbar^2} \right)^{-3/2} \left( \frac{2T_r}{T} \right). \tag{16}$$

Here $\mu$ is molar mass of gas, $T_r$ is the activation temperature for rotational degrees of freedom, and $P_0 = 1$ bar.

Deriving $\partial \Delta\nu / \partial \Delta H_{ads}$ for the RMT gives the values for optimal enthalpy $\Delta H_{ads}^{opt}$. Thus, the optimal enthalpy for the restricted multilayer case is given by eq. (17).

$$\Delta H_{ads}^{opt} = RT \ln \beta_1 = RT \ln \tilde{\beta}_1 / \alpha, \tag{17}$$

where

$$\tilde{\beta}_1 = \frac{a(1 - m\tilde{\beta}_2) - b\sqrt{m}(1 - \tilde{\beta}_2)}{b\sqrt{m}(1 - \tilde{\beta}_2^n) - am(1 - m^n\tilde{\beta}_2^n)},$$

$$a = \sqrt{1 - (n+1)\tilde{\beta}_2^n + n\tilde{\beta}_2^{n+1}},$$

$$b = \sqrt{1 - (n+1)m^n\tilde{\beta}_2^n + nm^{n+1}\tilde{\beta}_2^{n+1}}.$$

Particularly for the high-temperature region, (e.g. $T$=298 K), and the associated value of $\alpha$ (16) we get

$$\Delta H_{ads}^{opt} = RT \ln \beta_1 = RT \ln \left( \tilde{\beta}_1 T^{7/2} / \left( \alpha_0 T_0^{7/2} \right) \right), \tag{18}$$

where $\alpha_0 = 5.1 \cdot 10^{-6}$, $T_0 = 298$ K and

$$\tilde{\beta}_1 = \alpha_0 \left( T_0/T \right)^{7/2} \beta_1.$$

Source Code for the computation of the enthalpy of adsorption under RMT for high and low temperatures can be found in Section IX Codes 3 and 2 respectively.

**Optimal Linker Length and void volume** We derive an approach to define optimal linker length for a series of frameworks. The linker length influences the asymptotic value of the isotherm (when $P \to \infty$). This is because the pore volume $V_p$, surface area $S_p$ and fraction of void volume $\gamma$ depend on the linker length.

To get the optimal linker length we consider the general equation for uptake (Eq. (14)) and, using one type of cell ($M = 1$), we obtain (Eq. (12)),

$$m = \frac{\gamma \mu V}{N_A \pi r_w^2} \frac{S_p}{V_p} \frac{N}{B} (\tilde{n}).$$

We've considered the factor $(N/B)(\tilde{n})$ in a previous section. Now we consider the second one

$$\gamma \frac{S_p}{V_p}$$

since all the other terms are constant and do not depend on the geometrical characteristics of the framework. We assume that the framework has one characteristic length $a$, which is the length of the linker. We assume that the net can be expanded based on a reduced unit cell. Total volume is given as $V \sim a^3$, but for the volume of the framework itself we suggest $V_{net} \sim a^{1+\xi}$, where $\xi$ indicates how the volume of the net grows with the growth of $a$. For instance, if $\xi = 0$, it means that the volume of the net grows only with linker length, whereas the cross section of linkers is constant.

We get

$$\gamma = 1 - \frac{V_{net}}{V} = 1 - \frac{\kappa}{a^{2-\xi}}, \tag{19}$$

where $\kappa$ is some constant depending on the framework. We also get $S_p/V_p = c/a$, where $c$ is constant for every framework not changing when enlarging the linkers. Thus,

$$f(a) = \gamma \frac{S_p}{V_p} = \left( 1 - \frac{\kappa}{a^{2-\xi}} \right) \frac{c}{a} \to \text{Maximize } f(a).$$

This gives

$$a^{2-\xi} = (3 - \xi)\kappa$$

and therefore

$$a = ((3-\xi)\kappa)^{\frac{1}{2-\xi}}. \tag{20}$$

The factor $\kappa$ can be defined from experimental data if $\gamma$ is known. Therefore, we obtain the universal recipe of best linker length under these assumptions.

For the high temperature region (for instance, for $T = 298$ K), we should consider not only translational part of the chemical potential for hydrogen, but also its rotational part (as the rotational activation temperature for hydrogen molecule is $T_r = 85.4$ K). The appropriate considerations (see section VI and VII) show that considerations of the full chemical potential in our theory only alter $\alpha$ as follows (see eq. (21))

$$\alpha \to \alpha(2 T_r)/T, \tag{21}$$

which for $T = 298$ K gives $\alpha = 5.1 \cdot 10^{-6}$ for full chemical potential. This can be compared with the value obtained at 77 K, $\alpha = 2.6 \cdot 10^{-4}$. Thus, all the equations from previous sections are valid for the high temperature region where the appropriate substitution of $\alpha$ is made.

## II.  ISOTHERMS FOR DIFFERENT $\Delta H_{ads}^*$ IN THE RESTRICTED MULTILAYER CASE

Let us plot isotherms for values of $\Delta H_{ads}^*$ different from 0.904kJ/mol. Isotherms for $\Delta H_{ads}^* = 0$ and $\Delta H_{ads}^* = 0.5$ kJ/mol are similar (see Figure 6 and Figure 7) whereas the ones for $H_{ads}^* = 2$ kJ/mol (see Figure 8) are different due to the significant impact of higher levels.

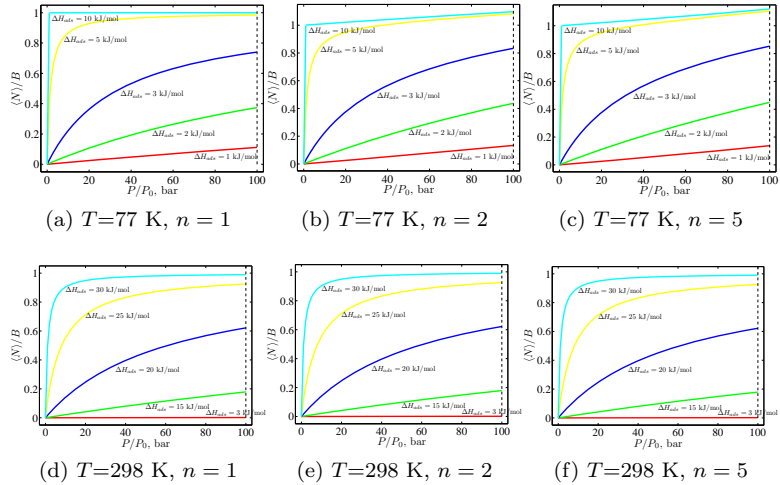

(a) $T$=77 K, $n = 1$    (b) $T$=77 K, $n = 2$    (c) $T$=77 K, $n = 5$

(d) $T$=298 K, $n=1$    (e) $T$=298 K, $n = 2$    (f) $T$=298 K, $n = 5$

FIG. 5: Isotherms for $H_2$ in the restricted multilayer case when $\Delta H_{ads}^* = 0.904$ kJ/mol at $T$=77 K and $T$=298 K.

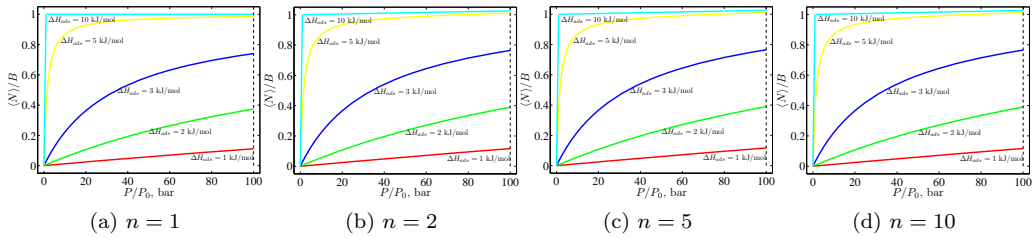

(a) $n = 1$    (b) $n = 2$    (c) $n = 5$    (d) $n = 10$

FIG. 6: Isotherms for $H_2$ in restricted multilayer case ($n = 1, 2, 5, 10$) when $\Delta H_{ads}^* = 0$ kJ/mol for $T$=77 K.

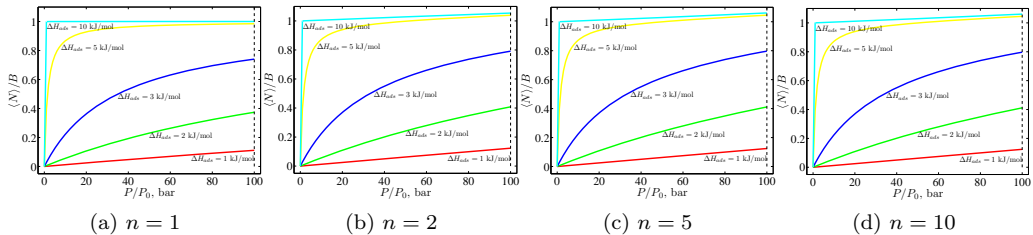

(a) $n = 1$    (b) $n = 2$    (c) $n = 5$    (d) $n = 10$

FIG. 7: Isotherms for $H_2$ in restricted multilayer case ($n = 1, 2, 5, 10$) when $\Delta H_{ads}^* = 0.5$ kJ/mol for $T$=77 K.

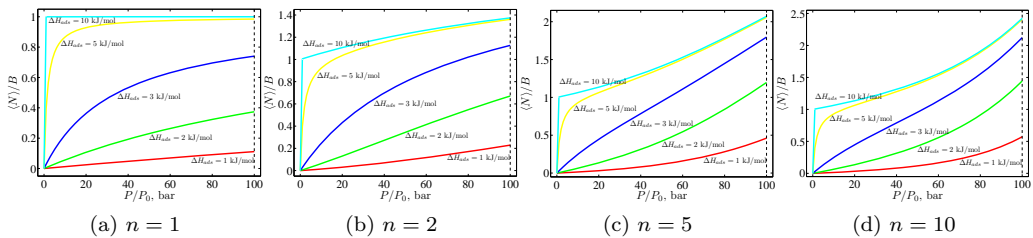

(a) $n = 1$    (b) $n = 2$    (c) $n = 5$    (d) $n = 10$

FIG. 8: Isotherms for $H_2$ in restricted multilayer case ($n = 1, 2, 5, 10$) when $\Delta H_{ads}^* = 2$ kJ/mol for $T$=77 K.

## A. Optimal parameters of the framework for the highest gas delivery using the Langmuir and BET theory

**Enthalpy of adsorption $\Delta H_{ads}$.** We define

$$\beta_1 = \exp\{\Delta H_{ads}/(RT)\}, \ \beta_2 = \exp\{\Delta H_{ads}^*/(RT)\},$$

$$\tilde{\beta}_1 = \alpha\beta_1, \ \tilde{\beta}_2 = \alpha\beta_2,$$

$$c = \beta_1/\beta_2, \ x = \alpha P/P_0\beta_2,$$

where $\alpha$ is a function of temperature that was defined previously. Deriving $\partial\Delta\nu/\partial\Delta H_{ads}$ for monolayer and multilayer theories gives the following values for optimal enthalpy $\Delta H_{ads}^{opt}$.

For the monolayer case, we get the optimal value of adsorption enthalpy $\Delta H_{ads}$ as follows.

$$\Delta H_{ads}^{opt} = RT\ln\frac{1}{\alpha\sqrt{m}} = RT\ln\left(\frac{T^{5/2}}{\alpha_0 T_0^{5/2}}\sqrt{\frac{P_0}{P}}\right) \quad (22)$$

where $\alpha_0 = 2.6$, and $\cdot 10^{-4}$, $T_0 = 77 \ K$. For instance, for pressure $P$=100 bar and $T$=77 K we get $\Delta H_{ads} = 3.75$ kJ/mol.

To obtain the optimal enthalpy in the monolayer case for the high-temperature region (for instance, for $T$=298 K) we make the appropriate change of $\alpha$ (see eq. (21)) and get

$$\Delta H_{ads}^{opt} = RT\ln\frac{1}{\alpha\sqrt{m}} = RT\ln\left(\frac{T^{7/2}}{\alpha_0 T_0^{7/2}}\sqrt{\frac{P_0}{P}}\right), \quad (23)$$

where $\alpha_0 = 5.1\cdot10^{-6}$ and $T_0 = 298$ K. Evaluating (23) for pressure $P$=100 bar and $T$=298 K, we get $\Delta H_{ads} = 24.5$ kJ/mol.

To calculate $\Delta\nu$ in the monolayer case using the calculated optimal adsorption enthalpy, $\Delta H_{ads}^{opt}$, we obtain,

$$\Delta\nu = \frac{\sqrt{m}-1}{\sqrt{m}+1}. \quad (24)$$

Substituting $P$=100 bar into (24), we get

$$\Delta\nu = 9/11 = 81.8\%.$$

The optimal adsorption enthalpy in **multilayer case** is given by eq. (25), given as

$$\Delta H_{ads}^{opt} = RT\ln\frac{1}{\alpha\sqrt{m}} + \beta_2 =$$
$$RT\ln\left(\frac{T^{5/2}}{\alpha_0 T_0^{5/2}}\sqrt{\frac{P_0}{P}} + \exp\frac{\Delta H_{ads}^*}{RT}\right). \quad (25)$$

Comparing (22) and (25) indicates that as $T \to \infty$ (and even at relatively high temperature), there is nearly no difference in the optimal enthalpy of adsorption between the monolayer and multilayer cases. For instance, for $P$=100 bar and $T$=77 K, this difference is only 8 J/mol. But when $T \to 0$ in the monolayer case, $\Delta H_{ads}^{opt} \to 0$, whereas in the multilayer case, $\Delta H_{ads}^{opt} \to \Delta H_{ads}^*$.

To obtain the optimal enthalpy in the multilayer case for high-temperature regions (for instance, $T$=298 K), we make the appropriate change of $\alpha$ (see eq. (21)) and get

$$\Delta H_{ads}^{opt} = RT\ln\frac{1}{\alpha\sqrt{m}} + \beta_2 =$$
$$RT\ln\left(\frac{T^{7/2}}{\alpha_0 T_0^{7/2}}\sqrt{\frac{P_0}{P}} + \exp\frac{\Delta H_{ads}^*}{RT}\right), \quad (26)$$

where $\alpha_0 = 5.1\cdot10^{-6}$ and $T_0 = 298$ K.

Source code for the computation of the enthalpy of adsorption for monolayer (Langmuir) and multilayer (BET) theory at low and high temperatures is found in Section IX Codes 4 and 5 respectively.

To determine the difference in the factor $\nu$ between pressures $P_0$ and $P$, we get, in the multilayer case,

$$\Delta\nu = \frac{1+\alpha\beta_2\sqrt{m}}{1+\sqrt{m}}\left[\frac{\sqrt{m}}{1-\alpha m\beta_2} - \frac{1}{1-\alpha\beta_2}\right]. \quad (27)$$

As the assumption of an infinite number of layers is not realistic, from (27), we obtain a restriction on the range of pressures. This theory works if

$$m < \frac{1}{\alpha\beta_2}. \quad (28)$$

For $T$=298 K, the upper border according to (28) equals 137,200 bar, which is far more than the value of 100 bar considered previously. For $T$=77 K, the upper bound is also far more than that of 100 bar, and equals 955 bar. At 45.5 K the upper bound reaches P=100 bar and hence the multilayer theory is not applicable for this range of pressures and temperatures below 45.5 K.

From equations (22), (25) and (17) we determine that optimal enthalpy increases with temperature.

We plot in color the dependence of the optimal enthalpy on temperature, $T$, and pressure, $P$, in all three cases for low and high temperatures: monolayer (see Figure 9a, d), multilayer (see Figure 9b, e) and restricted multilayer for 3 layers (see Figure 9c, f). For molecular hydrogen, $\Delta H_{ads}^*$ is relatively low. Hence the graphs shown in Figure 9d,e,f for different number of layers are identical. Discrepancies are significant only at very low temperatures.

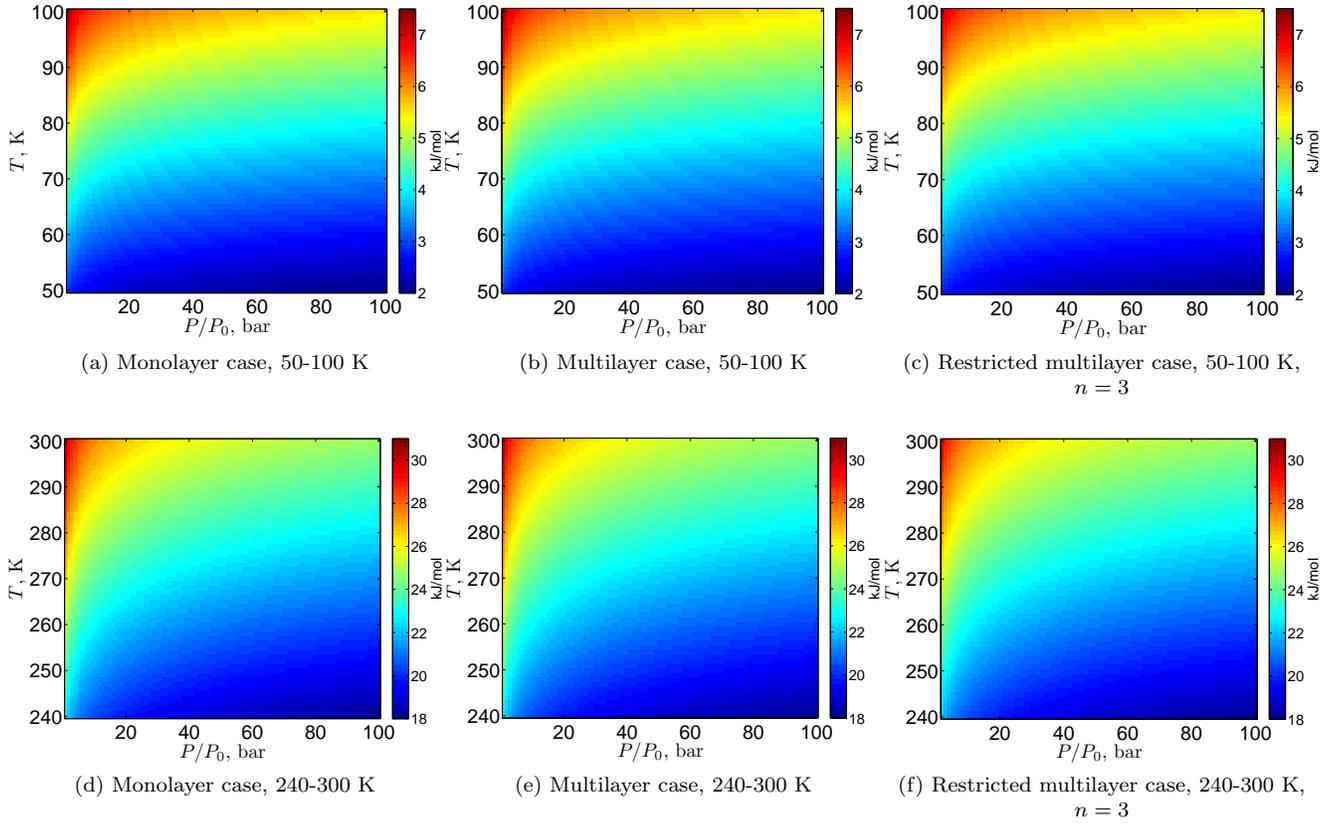

(a) Monolayer case, 50-100 K

(b) Multilayer case, 50-100 K

(c) Restricted multilayer case, 50-100 K, $n = 3$

(d) Monolayer case, 240-300 K

(e) Multilayer case, 240-300 K

(f) Restricted multilayer case, 240-300 K, $n = 3$

FIG. 9: Temperature and pressure dependence of optimal enthalpy $\Delta H_{ads}^{opt}$ for low and high temperature region

## III. CUBIC FRAMEWORK WITH INTERPENETRATION.

Another framework explored in this work is one where there is interpenetration present, where frameworks of the same kind are catenated starting with a two-fold degree and eventually reaching a maximum limited by the space inside the pores.

We consider the framework of $q$ interpenetrated cubic nets with linkers that are cylinders of length $l$ and base radii $r$, and SBU-nodes that we treat as spheres with diameter $d$. If the volume of the porous material is $V$, then we get

$$\frac{V}{(d+l)^3}$$

cells and therefore the uptake in number of hydrogen molecules would be

$$N = \frac{V}{(d+l)^3} B \frac{N}{B} (\tilde{n}),$$

where $\tilde{n}$ is an effective number of layers, $B$ is a number of sites and $\frac{N}{B}(\tilde{n})$ is $\frac{N}{B}$ for restricted multilayer case with $\tilde{n}$ layers. In order to estimate the effective number of layers one can consider the longest distance in a cubic cell, which is its main diagonal, and thus get the effective number of layers in the following equation,

$$\tilde{n} = \frac{\sqrt{3}(d+l)}{D+\delta},$$

where, $D$ is the diameter of a hydrogen molecule and $\delta$ is the characteristic distance between them. Another approach takes into account the symmetry of adsorption in a cubic cell and therefore the effective number of layers is smaller. In this approach the effective longest distance is not the diagonal but is instead the edge of a cube. Therefore, we get

$$\tilde{n} = \frac{d+l}{D+\delta}. \tag{29}$$

Thus we use eq. (29) for estimating the effective number of layers. We should also note that if eq. (29) gives fractional number of layers e.g. 1.54, one can pick up either 1 or 2 layers. In this article we use the biggest integer number not higher than $\tilde{n}$.

Let us consider the case of cubic frameworks with an interpenetration of degree $q$. In this case, the length of the linker is $l$, the diameter of the SBU-node, which is considered to be a sphere, is $d$, and the radius of the linker is $r$. The factor $\gamma$, which, in this case, defines the share of the free volume, equals

$$\gamma = 1 - q\frac{\pi/6d^3 + 3\pi r^2 l}{(d+l)^3}.$$

The surface area of a pore is equal to

$$S_p = 6(d+l)^2 - 6\pi d^2/4 = 6\left[(d+l)^2 - \pi d^2/4\right].$$

Given that the degree of interpenetration is $q$, we can say that

$$S = qS_p,$$

and therefore

$$B = \frac{qS_p}{\pi r_{H_2}^2}.$$

Now let us apply this equation to determining the adsorption in MOF-9, which is a cubic interpenetrated framework of degree two. For this MOF, $d = 9.65\,\text{Å}$, $l = 11.88\,\text{Å}$ and $2r = 5\,\text{Å}$. Thus we get

$$\gamma = 1 - 2 \cdot 0.119 = 0.762.$$

This agrees well with experimental measurements of $\gamma = 0.71$ that were carried out by Eddaoudi et al. [4]. According to (30), $S_p = 2340\,\text{Å}^2$, and therefore the number of sorption sites is $B = 194.16 \approx 194$, and the number of layers is $\tilde{n} = 3.9 \approx 4$. The uptake (Eq. 12) gives

$$m = 45.9\frac{N}{B}(4)\ \text{g/L}.$$

## IV. COMPARISON TO EXPERIMENTS

### A. MOF-5

We assume that the discrepancy in the low-pressure region can also be caused a by slight decrease of the adsorption enthalpy reported in [5]. This assumption is further supported by the fact that excess uptake in the region of high pressures and low temperatures is mainly sensitive not to changes in adsorption enthalpy, but to changes in pore properties [6]. We will use two optimized enthalpy estimations. The first is the sum of squares of deviations of uptake. The second is the sum of the absolute values of deviations. The dependency of these parameters on enthalpy of adsorption can be seen in Figure 10a and Figure 10b.

The optimal values of $\Delta H_{ads}$ are very close: 4.47 kJ/mol and 4.43 kJ/mol, respectively. Isotherms with these enthalpies of adsorption are plotted on the Figure 11a and Figure 11b. The square root of the average square deviation and absolute deviation are, respectively, 0.81 (g/L) and 0.65 g/L.

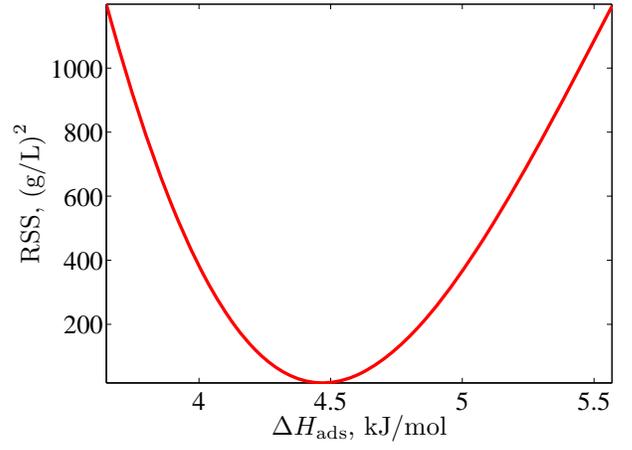

(a) Sum squares of deviations

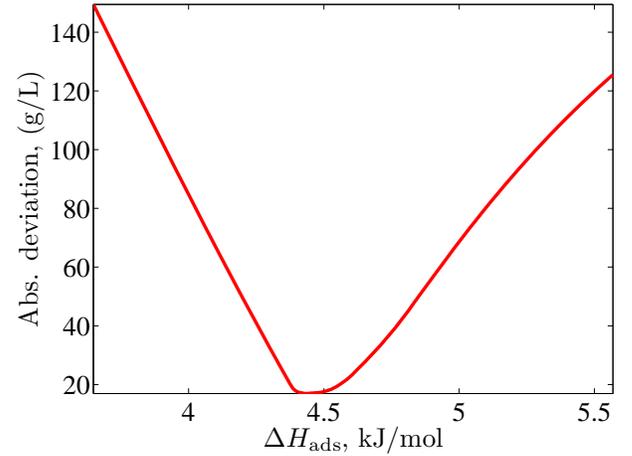

(b) Sum absolute values of deviations

FIG. 10: Two measures of fit for MOF-5

### B. MOF-177

We assume that it is caused by the slight decrease of adsorption enthalpy reported in [5] which can also be supported by the fact that excess uptake in region of high pressures and low temperatures is mainly sensitive not to changes in adsorption enthalpy, but to changes in pore properties [6]. We will use the same two optimized enthalpy estimations as was done in the previous section. The dependency of these parameters on enthalpy of adsorption can be seen in Figure 12 a and Figure 12b.

The optimal values of $\Delta H_{ads}$ are very close: 4.07 kJ/mol and 4.05 kJ/mol, respectively. Isotherms for these enthalpies of adsorption are displayed in Figure 13a and Figure 13b. The square root of the average square deviation and the absolute deviation are, respectively, 0.64 (g/L) and 0.51 g/L.

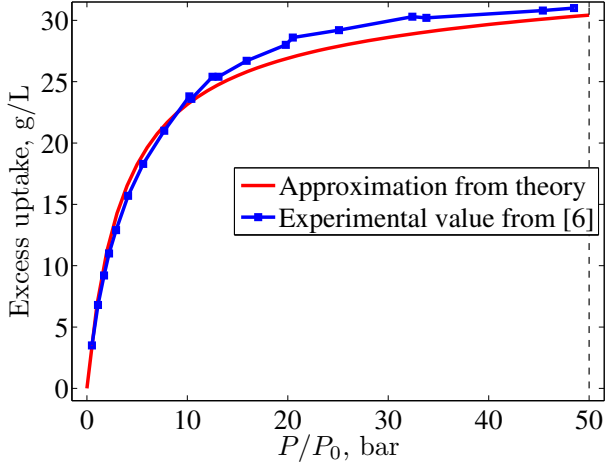

(a) $\Delta H_{ads} = 4.470$ kJ/mol

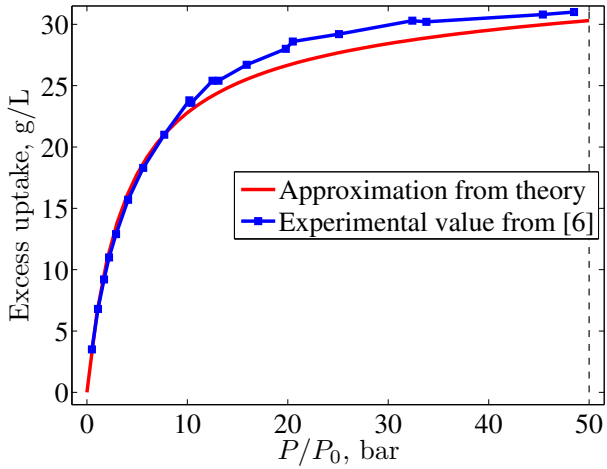

(b) $\Delta H_{ads} = 4.431$ kJ/mol

FIG. 11: Isotherm for MOF-5 for two obtained optimal enthalpies $\Delta H_{ads}^*$

## V. EFFECTIVE RADIUS OF HYDROGEN

Now we calculate the effective radius of hydrogen necessary for the number of sites calculations. Let the relation between the number of available sites and the surface area be

$$B = \theta_0 S.$$

### A. T=77 K

To obtain the coefficient we use the data from [5],

$$N = \frac{S}{S_0} = B\frac{N}{B}(\tilde{n}) = \theta_0 S \frac{N}{B}(\tilde{n}),$$

where $S_0$ is the characteristic surface area occupied by one adsorbed molecule of hydrogen and S is a surface

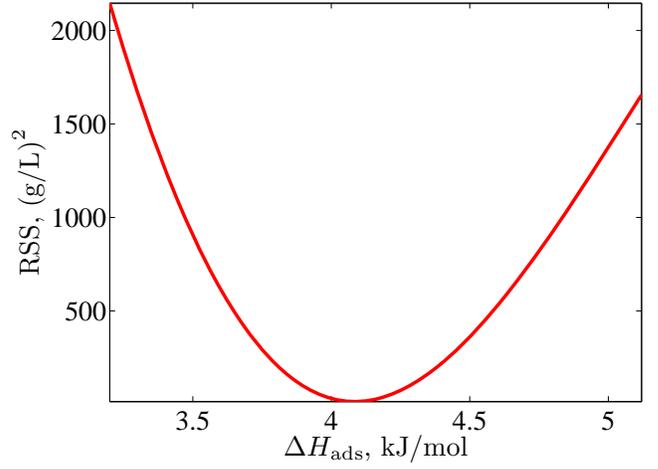

(a) Sum squares of deviations

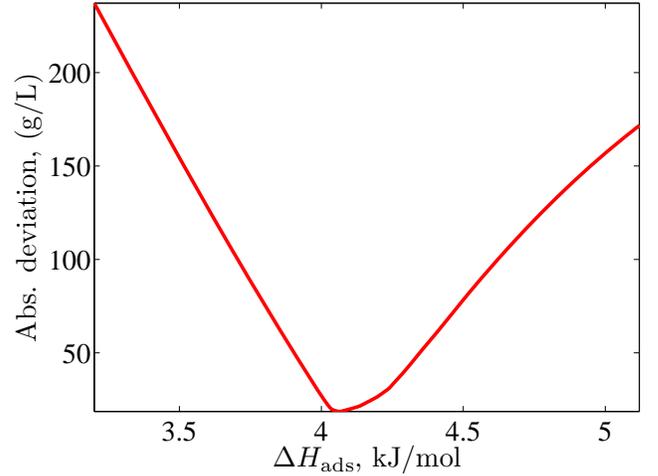

(b) Sum absolute values of deviations

FIG. 12: Two measures of fit for MOF-177

TABLE I: Surface area and excess uptake for different MOFs

| Surface area, m²/g | Uptake, mg/g | MOF |
|---|---|---|
| 1100 | 22 | MOF-74 |
| 2300 | 34 | HKUST-1 |
| 2400 | 37 | IRMOF-11 |
| 3300 | 50 | IRMOF-6 |
| 4200 | 53 | IRMOF-1 |
| 4700 | 68 | IRMOF-20 |
| 5700 | 75 | MOF-177 |

area. We will make the assumption that

$$\frac{N}{B}(\tilde{n})$$

does not depend on MOF, which is based on the insignificant impact of slight changes in the enthalpy of adsorption at such a low temperature. The other way to count

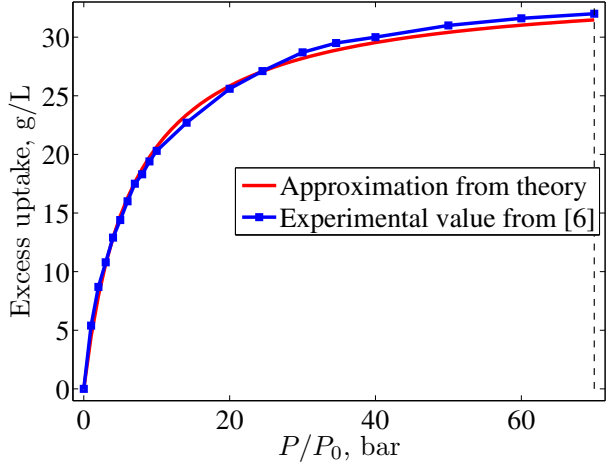

(a) $\Delta H_{ads} = 4.067$ kJ/mol

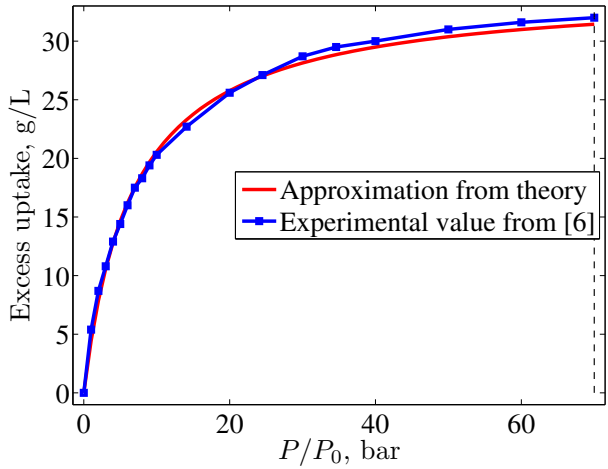

(b) $\Delta H_{ads} = 4.054$ kJ/mol

FIG. 13: Isotherm for MOF-177 for two obtained optimal enthalpies $\Delta H_{ads}^*$

the excess uptake is as follows,

$$\theta = \frac{m}{S} = \frac{N\mu}{N_A S} = \frac{2}{N_A S_0}\,\text{g/m}^2.$$

When fitting a linear model for data from table SII, we obtain

$$\theta = 0.0139 \text{mg/m}^2.$$

Therefore, we get

$$S_0 = 10^{20}\text{Å}^2\,\frac{2\,\text{g/mol}}{0.0139\cdot 10^{-3}\text{g}\,6.02\cdot 10^{23}\text{mol}^{-1}} = 23.9\,\text{Å}^2.$$

And thus the effective radius of hydrogen equals

$$r_{H_2} = \sqrt{\frac{S_0}{\pi}} = 2.76\,\text{Å}.$$

## B.  T=298 K

To obtain the coefficient we use the data from [7],

TABLE II: Surface area and excess uptake for different Li-MOFs

| Surface area, $\text{m}^2$/g | Uptake, mg/g | MOF |
|---|---|---|
| 3047 | 50.86 | IRMOF-1-4Li |
| 3620 | 58.97 | IRMOF-8-6Li |
| 4135 | 72.77 | MOF-177-15Li |
| 5558 | 87.34 | MOF-200-27Li |

$$N = \frac{S}{S_0} = B\frac{N}{B}\,(\tilde{n}) = \theta_0 S \frac{N}{B}\,(\tilde{n})\,,$$

where $S_0$ is characteristic surface area occupied by one adsorbed molecule of hydrogen and S is a surface area. We will make the assumption that

$$\frac{N}{B}\,(\tilde{n})$$

does not depend on MOF because, at such a pressure and such a high enthalpy of adsorption, the saturation of the first layer is nearly reached, whereas the second and subsequent layers are not as attracted to one another, since $\Delta H_{ads}^* < RT$. The other way to count the excess uptake is the following

$$\theta = \frac{m}{S} = \frac{N\mu}{N_A S} = \frac{2}{N_A S_0}\,\text{g/m}^2.$$

After fitting a linear model for the data from table SII, we obtain

$$\theta = 0.0164 \text{mg/m}^2.$$

Therefore we get

$$S_0 = 10^{20}\text{Å}^2\,\frac{2\,\text{g/mol}}{0.0139\cdot 10^{-3}\text{g}\,6.02\cdot 10^{23}\text{mol}^{-1}} = 20.3\,\text{Å}^2.$$

And thus the effective radius of hydrogen equals

$$r_{H_2} = \sqrt{\frac{S_0}{\pi}} = 2.54\,\text{Å}.$$

We will also use this effective radius of hydrogen for temperatures which are close to 298 K, for 243 and 273 K.

## VI.  CHEMICAL POTENTIAL FOR HIGH TEMPERATURES

For low temperatures (e.g., for 77 K), rotations in molecular hydrogen are not significant, because the rotation activation temperature is 85.4 K and therefore we only used the translational part of chemical potential. Here we derive the rotational chemical potential for high temperatures. Consideration of both high temperature and low temperature cases can be found in [8]. As the vibration activation temperature is 6100 K, we do not consider the vibrational part of chemical potential.

Due to nuclear symmetry, the normal term of a hydrogen molecule has different degree of degeneration for rotational levels with odd and even orbital moment $K$. The molecules with an even total spin ($S = 0$ for hydrogen) must have an even orbital moment $(0, 2, \dots)$, whereas molecules with an odd total spin ($S = 1$ for hydrogen) must have an odd orbital moment $(1, 3, \dots)$. The energy corresponding to the orbital movement with moment $K$ is

$$\frac{\hbar^2 K(K+1)}{2I},$$

where $I = m/2\, r_0^2$, where $m$ is a mass of hydrogen atom and $r_0$ is the equilibrium distance between atoms in molecule. We introduce the designation

$$kT_r = \hbar^2/2I,$$

where $T_r$ is the rotation activation temperature. For molecular hydrogen, $T_r = 85.4$ K. Further hydrogen modification with odd spin is called ortho- or u-hydrogen, whereas hydrogen modification with even spin is called para- or g-hydrogen. Relative degenerations of these two modifications are $g_u = 3/4$ and $g_g = 1/4$, respectively. Thus the partition functions for the modifications are

$$Z_u = \sum_{K=1,\,3,\dots} (2K+1)\exp(-T_r/TK(K+1)), \quad (30)$$

and

$$Z_g = \sum_{K=0,\,2,\dots} (2K+1)\exp(-T_r/TK(K+1)). \quad (31)$$

Therefore the full partition function is

$$Z = g_u Z_u + g_g Z_g = \frac{Z_g + 3Z_u}{4}. \quad (32)$$

For high temperatures,

$$Z_u \approx Z_g \approx 1/2 \sum_{K=0,\,1,\,2,\dots} (2K+1)\exp(-T_r/TK(K+1))$$

which we approximate by an integral

$$\sum_{K=0,\,1,\,2,\dots} (2K+1)\exp(-T_r/TK(K+1)) \approx$$

$$\sum_{K=0,\,1,\,2,\dots} 2K\exp(-T_r/TK^2) \approx$$

$$\int_0^\infty \exp(-T_r/TK^2)dK^2 = T/T_r.$$

Thus $Z \approx T/(2\,T_r)$ and the corresponding free energy

$$F_{rot} = -NkT \log \frac{T}{2\,T_r}.$$

Finally we get the equation for rotational part of chemical potential for high temperatures

$$\mu_{rot} = -kT \log \frac{T}{2\,T_r}. \quad (33)$$

## VII.   THEORY WHEN CONSIDERING FULL CHEMICAL POTENTIAL

In this section we show how it is necessary to alter the calculations which were previously presented for the translational part of chemical potential in order to consider the rotational part of the chemical potential. We use $\lambda = \exp(\mu/RT)$.

$$\lambda = \exp(\mu/kT) = \exp(\mu_{tr}/kT) \times \exp(\mu_{rot}/kT) = \\ \alpha P/P_0 \exp(\mu_{rot}/kT), \quad (34)$$

where $\mu_{tr}$ and $\mu_{rot}$ are the translational and rotational parts of chemical potential, respectively. As the enthalpy of adsorption is relatively high, we can still consider adsorption as localized. Therefore, from eq. (34) and eq. (33), we get

$$\lambda = \alpha \frac{2T_r}{T} P/P_0. \quad (35)$$

From equation (35), we obtain that the consideration of full chemical potential results only in an appropriate change of $\alpha$

$$\alpha_{new} = \alpha \frac{2T_r}{T}. \quad (36)$$

## VIII.   DEPENDENCE OF ENTHALPY OF ADSORPTION ON PRESSURE

For Li-metalated frameworks with high enthalpy of adsorption that face high uptake at room temperature, there is a significant dependence of the enthalpy of adsorption on uptake and therefore on pressure. Here we illustrate how this dependence was calculated for MOF-200-27Li at room temperature. At the end of this section we provide tables for calculated enthalpies for MOF200-27Li and MOF-177-15Li at 243, 273 and 298 K.

From [7], we get the dependence of adsorption enthalpy on total weight uptake, for which we use the following linear approximation

$$\Delta H_{ads} = Q_{st} = (29 - u)\,\text{kJ/mol},$$

where $u$ is the weight total uptake in units of percent. As $\Delta H_{ads}^* \ll RT \ll \Delta H_{ads}$, we use monolayer theory. We also make an assumption that reaching 10% of total uptake is nearly the same as total uptake at saturation. We also apply our theoretical approximation of excess uptake to total uptake, and therefore we obtain

$$u = \frac{10}{1 + 1/\alpha \cdot P_0/P \exp(-\Delta H_{ads}/RT)}.$$

Thus we get the following equation

$$29 - \Delta H_{ads}/1000 = \frac{10}{1 + 1/\alpha \cdot P_0/P \exp(-\Delta H_{ads}/RT)}. \quad (37)$$

For $T = 298$ K, equation (37) gives

$$29 - 2.4764\,x = \frac{10}{1 + 1.954 \cdot 10^5/m \cdot e^{-x}},$$

where $x = \Delta H_{ads}/RT$ and $m = P/P_0$. Solving the last equation at different pressures $P$, we obtain the data for enthalpy at $T = 298$ K at different pressures (see tab. III). Using the same approach we obtain the enthalpies for other temperatures.

TABLE III: Enthalpy of adsorption of MOF-200-27Li for different pressures for T=243, 273 and 298 K

| $P$, bar | $\Delta H_{ads}/RT$ for $T$=298 K | $\Delta H_{ads}/RT$ for $T$=273 K | $\Delta H_{ads}/RT$ for $T$=243 K |
|---|---|---|---|
| 1 | 10.8607 | 11.2894 | 11.7611 |
| 5 | 10.1310 | 10.4301 | 10.8103 |
| 10 | 9.7863 | 10.0511 | 10.4133 |
| 20 | 9.4388 | 9.6790 | 10.0346 |
| 30 | 9.2383 | 9.4681 | 9.8251 |
| 40 | 9.0988 | 9.3228 | 9.6831 |
| 50 | 8.9927 | 9.2131 | 9.5774 |
| 60 | 8.9078 | 9.1258 | 9.4942 |
| 70 | 8.8372 | 9.0536 | 9.4261 |
| 80 | 8.7772 | 8.9925 | 9.3690 |
| 90 | 8.7252 | 8.9397 | 9.3201 |
| 100 | 8.6795 | 8.8934 | 9.2776 |

TABLE IV: Enthalpy of adsorption of MOF-177-15Li for different pressures for T=243, 273 and 298 K

| $P$, bar | $\Delta H_{ads}/RT$ for $T$=298 K | $\Delta H_{ads}/RT$ for $T$=273 K | $\Delta H_{ads}/RT$ for $T$=243 K |
|---|---|---|---|
| 1 | 10.4573 | 10.9922 | 11.5831 |
| 5 | 9.9044 | 10.2709 | 10.7403 |
| 10 | 9.6198 | 9.9404 | 10.3879 |
| 20 | 9.3265 | 9.6143 | 10.0561 |
| 30 | 9.1562 | 9.4303 | 9.8760 |
| 40 | 9.0378 | 9.3044 | 9.7560 |
| 50 | 8.9481 | 9.2100 | 9.6680 |
| 60 | 8.8765 | 9.1354 | 9.5997 |
| 70 | 8.8174 | 9.0742 | 9.5445 |
| 80 | 8.7674 | 9.0227 | 9.4988 |
| 90 | 8.7242 | 8.9785 | 9.4601 |
| 100 | 8.6864 | 8.9400 | 9.4268 |

# IX.  SOURCE CODES FOR MATLAB

### Code 1: Occupancy Ratio Calculation

```matlab
% Calculates occupancy ratio (N / B) given a temperature, number of layers.
% Enthalpy of adsorption for higher layers is set to that of hydrogen.
% The eqaltion holds for low temperature. To make calculation for high
% temperature one should make a proper adjustment for alpha (see Supp. Info)

T=77;
T_0=298;
R=8.31;
H_star=904;

alpha=8.9e-6*power(T_0/T,2.5);
beta=H_star/R/T;

n=3;

H_ads=4400;
par=H_ads/R/T;
p=70;
x=alpha*exp(beta)*p;

c=exp(par-beta);
n_total=(c*x*(1-(n+1)*power(x,n)+n*power(x,n+1))/(1-x+c*x-c*power(x,n+1)))/(1-x)
```

### Code 2: Optimal Enthalpy Calculation for Restricted Multilayer Theory at High Temperatures

```matlab
% Plots the optimal enthalpy in given range of temperatures for restricted
% multilayer case for specified number of layers for low
% temperature region. Since the optimal enthalpy varies only slightly for
% different number of layers the results are very exact for any restricted
% multilayer case as well.

n=6;

T_min=50;
T_max=100;
T=linspace(T_min,T_max,T_max-T_min+1);
T_0=298;
R=8.31;
alpha_0=8.9e-6;
H_star=904;

m=linspace(2,100,99);
H_opt_restr_multi=zeros(size(T,2),size(m,2));

for t=1:size(T,2)
    for mul=1:size(m,2)
        alpha=alpha_0*power(T_0/T(1,t),2.5);
        beta2=exp(H_star/R/T(1,t));

        beta2_tilde=alpha*beta2;

        a=sqrt(1-(n+1)*power(beta2_tilde,n)+n*power(beta2_tilde,n+1));
        b=sqrt(1-(n+1)*power(m(1,mul)*beta2_tilde,n)+n*power(m(1,mul)*beta2_tilde,n+1));

        beta1_tilde=(a*(1-m(1,mul)*beta2_tilde)-(1-beta2_tilde)*b*sqrt(m(1,mul)))/...
        (b*sqrt(m(1,mul))*(1-power(beta2_tilde,n))-m(1,mul)*a*(1-power(m(1,mul)*beta2_tilde,n)));

        beta1=beta1_tilde/alpha;

        H_opt_restr_multi(t,mul)=R*T(1,t)*log(beta1);
    end
end

h=figure;

mu=imagesc(m',T',H_opt_restr_multi/1000);

xlabel('$P/P_0$, bar','Interpreter','latex','FontSize',24);
ylabel('$T$, K','Interpreter','latex','FontSize',24);

set(gca,'FontSize',24);
set(gca,'Position',[.12 .15 .8 .8])
set(gca,'YDir','normal');
g=colorbar('location','eastoutside','FontSize',18);
%caxis([18 31]);
caxis([2 7.5]);
ylabel(g,'kJ/mol');

str=strcat('H_opt_restr_multilayer',int2str(T_min),'_',int2str(T_max),'_',int2str(n));

saveas(h,strcat(str,'.png'),'png');
saveas(h,strcat(str,'.eps'),'psc2');
```

## Code 3: Optimal Enthalpy Calculation for Restricted Multilayer Theory at Low Temperatures

```matlab
% Plots the optimal enthalpy in given range of temperatures for restricted
% multilayer case for specified number of layers for low
% temperature region. Since the optimal enthalpy varies only slightly for
% different number of layers the results are very exact for any restricted
% multilayer case as well.

n=6;

T_min=50;
T_max=100;
T=linspace(T_min,T_max,T_max-T_min+1);
T_0=298;
R=8.31;
alpha_0=8.9e-6;
H_star=904;

m=linspace(2,100,99);
H_opt_restr_multi=zeros(size(T,2),size(m,2));

for t=1:size(T,2)
    for mul=1:size(m,2)
        alpha=alpha_0*power(T_0/T(1,t),2.5);
        beta2=exp(H_star/R/T(1,t));

        beta2_tilde=alpha*beta2;

        a=sqrt(1-(n+1)*power(beta2_tilde,n)+n*power(beta2_tilde,n+1));
        b=sqrt(1-(n+1)*power(m(1,mul)*beta2_tilde,n)+n*power(m(1,mul)*beta2_tilde,n+1));

        beta1_tilde=(a*(1-m(1,mul)*beta2_tilde)-(1-beta2_tilde)*b*sqrt(m(1,mul)))/...
        (b*sqrt(m(1,mul))*(1-power(beta2_tilde,n))-m(1,mul)*a*(1-power(m(1,mul)*beta2_tilde,n)));

        beta1=beta1_tilde/alpha;

        H_opt_restr_multi(t,mul)=R*T(1,t)*log(beta1);
    end
end

h=figure;

mu=imagesc(m',T',H_opt_restr_multi/1000);

xlabel('$P/P_0$, bar','Interpreter','latex','FontSize',24);
ylabel('$T$, K','Interpreter','latex','FontSize',24);

set(gca,'FontSize',24);
set(gca,'Position',[.12 .15 .8 .8])
set(gca,'YDir','normal');
g=colorbar('location','eastoutside','FontSize',18);
%caxis([18 31]);
caxis([2 7.5]);
ylabel(g,'kJ/mol');

str=strcat('H_opt_restr_multilayer',int2str(T_min),'_',int2str(T_max),'_', int2str(n));

saveas(h,strcat(str,'.png'),'png');
saveas(h,strcat(str,'.eps'),'psc2');
```

Code 4: Optimal Enthalpy Calculation for Langmuir and BET Theories at Low Temperatures

```matlab
% Plots the optimal enthalpy in given range of temperatures for both
% monolayer case (par = 0) and multilayer case (par = 1) for low
% temperature region. Since the optimal enthalpy varies only slightly for
% different number of layers the results are very exact for any restricted
% multilayer case as well.

T_0=298;
T_min=50;
T_max=100;
T=linspace(T_min,T_max,T_max-T_min+1);
R=8.31;
alpha_0=8.9e-6;
par=0; %equals 1 when it's multilayer case
H_star=904;

m=linspace(1,100,100);
H_opt_mono=zeros(size(T,2),size(m,2));
H_opt_multi=zeros(size(T,2),size(m,2));

for t=1:size(T,2)
    for mul=1:size(m,2)
        H_opt_mono(t,mul)=R*T(1,t)*log(power(T(1,t)/T_0,2.5)...
            /sqrt(m(1,mul))/alpha_0);
        H_opt_multi(t,mul)=R*T(1,t)*log(power(T(1,t)/T_0,2.5)...
            /sqrt(m(1,mul))/alpha_0+exp(H_star*R/T(1,t)));
    end
end
h=figure;

if (par==0)
    mu=imagesc(m',T',H_opt_mono/1000);
else
    mu=imagesc(m',T',H_opt_multi/1000);
end
xlabel('$P/P_0$, bar','Interpreter','latex','FontSize',24);
ylabel('$T$, K','Interpreter','latex','FontSize',24);

set(gca,'Position',[.12 .15 .8 .8])
set(gca,'FontSize',24);
set(gca,'YDir',' normal');
g=colorbar('location','eastoutside','FontSize',18);
%caxis([20,29])
caxis([2 7.5])

ylabel(g,'kJ/mol');

if (par==0)
    str=strcat('H_opt_monolayer_',int2str(T_min),'_',int2str(T_max));
else
    str=strcat('H_opt_multilayer_',int2str(T_min),'_',int2str(T_max));
end
saveas(h,strcat(str,'.png'),'png');
saveas(h,strcat(str,'.eps'),'psc2');

P_req=[2 10 30 100];

g=figure;

hold 'on'
plot(T,H_opt_mono(:,P_req(1,1))/1000,'g-','LineWidth',3);
plot(T,H_opt_mono(:,P_req(1,2))/1000,'r-','LineWidth',3);
plot(T,H_opt_mono(:,P_req(1,3))/1000,'b-','LineWidth',3);
plot(T,H_opt_mono(:,P_req(1,4))/1000,'c-','LineWidth',3);

%plot(T,H_opt_multi(:,P_req)/1000,'r-','LineWidth',2)

xlabel('T, K','Interpreter','latex','FontSize',24);
ylabel('$\Delta H_{\mathrm{ads}}^{\mathrm{opt}}$, kJ/mol','Interpreter','latex','FontSize',24);

legend('P=2 bar','P=10 bar','P=30 bar','P=100 bar');
set(legend,'FontSize',16,'FontName','Times','Location','Southeast');
set(gca,'FontSize',24);
set(gca,'Position',[.12 .15 .8 .8])
axis 'tight'
str=strcat('H_opt_diff_pressures');

saveas(g,strcat(str,'.png'),'png');
saveas(g,strcat(str,'.eps'),'psc2');
```

## Code 5: Optimal Enthalpy Calculation for Langmuir and BET Theories at High Temperatures

```matlab
% Plots the optimal enthalpy in given range of temperatures for both
% monolayer case (par = 0) and multilayer case (par = 1) for high
% temperature region. Since the optimal enthalpy varies only slightly for
% different number of layers the results are very exact for any restricted
% multilayer case as well.

T_0=298;
T_min=240;
T_max=300;
T=linspace(T_min,T_max,T_max-T_min+1);
R=8.31;
alpha_0=5.1e-6;
par=0; %equals 1 when it's multilayer case
H_star=904;

m=linspace(1,100,100);
H_opt_mono=zeros(size(T,2),size(m,2));
H_opt_multi=zeros(size(T,2),size(m,2));

for t=1:size(T,2)
    for mul=1:size(m,2)
        H_opt_mono(t,mul)=R*T(1,t)*log(power(T(1,t)/T_0,3.5)...
            /sqrt(m(1,mul))/alpha_0);
        H_opt_multi(t,mul)=R*T(1,t)*log(power(T(1,t)/T_0,3.5)...
            /sqrt(m(1,mul))/alpha_0+exp(H_star*R/T(1,t)));
    end
end
h=figure;

if (par==0)
    mu=imagesc(m',T',H_opt_mono/1000);
else
    mu=imagesc(m',T',H_opt_multi/1000);
end
xlabel('$P/P_0$, bar','Interpreter','latex','FontSize',24);
ylabel('$T$, K','Interpreter','latex','FontSize',24);

set(gca,'Position',[.12 .15 .8 .8])
set(gca,'FontSize',24);
set(gca,'YDir','normal');
g=colorbar('location','eastoutside','FontSize',18);
caxis([18,31])

ylabel(g,'kJ/mol');

if (par==0)
    str=strcat('H_opt_monolayer_high_temp',int2str(T_min),'_',int2str(T_max));
else
    str=strcat('H_opt_multilayer_high_temp',int2str(T_min),'_',int2str(T_max));
end
saveas(h,strcat(str,'.png'),'png');
saveas(h,strcat(str,'.eps'),'psc2');

P_req=[2 10 30 100];

g=figure;

hold 'on'
plot(T,H_opt_mono(:,P_req(1,1))/1000,'g-','LineWidth',3);
plot(T,H_opt_mono(:,P_req(1,2))/1000,'r-','LineWidth',3);
plot(T,H_opt_mono(:,P_req(1,3))/1000,'b-','LineWidth',3);
plot(T,H_opt_mono(:,P_req(1,4))/1000,'c-','LineWidth',3);

%plot(T,H_opt_multi(:,P_req)/1000,'r-','LineWidth',2);

xlabel('T, K','Interpreter','latex','FontSize',24);
ylabel('$\Delta H_{\mathrm{ads}}^{\mathrm{opt}}$, kJ/mol','Interpreter','latex','FontSize',24);

legend('P=2 bar','P=10 bar','P=30 bar','P=100 bar');
set(legend,'FontSize',16,'FontName','Times','Location','Southeast');
set(gca,'FontSize',24);
set(gca,'Position',[.12 .15 .8 .8])
axis 'tight'
str=strcat('H_opt_diff_pressures_high_temp');

saveas(g,strcat(str,'.png'),'png');
saveas(g,strcat(str,'.eps'),'psc2');
```

### Code 6: Optimal Parameters Calculation from Linker Length and Void Volume

```
1   % Calculates optimal parameters (see the model from the paper) of a
2   % framework given the linker lengths and void volumes for the series of
3   % equivalent frameworks. R_square is linear regression dtermination
4   % coefficient. The closer it is to 1 the better the model fits experimtal
5   % data.
6
7   function [kappa_opt xi_opt R_squared]=count_opt_par(gamma, a_approx)
8
9   x=log(a_approx);
10  y=log(1-gamma);
11  x=[ones(size(x,1),1) x];
12  res=inv(x'*x)*x'*y;
13  kappa_opt=exp(res(1,1));
14  xi_opt=2*res(2,1);
15
16
17  y_hat = x * res;
18  R_squared = 1 - (y - y_hat)' * (y - y_hat) / ((y - mean(y))' * (y - mean(y)));
19
20  return
```

### Code 7: Translational Chemical Potential Calculation

```
1   % Function implements calculation of translational chemical potential given
2   % temperature, n_q which is the function of gas molar mass and
3   % temperature and concentration (N / V)
4
5   function mu=count_mu(N, T, V, n_q)
6
7   k=1.38e-23;
8   hbar=1.05e-34;
9   mu=-k*T*log(2*V/N*n_q);
10
11  return
```

### Code 8: Calculation of Maximum Pressure where Restricted Multilayer Theory still holds

```
1   % Calculates and plots border pressure depending on temperature.
2   % The border pressure is a maximum pressure where the theory still holds.
3
4   T=linspace(40,300,261);
5   alpha0=8.9e-6;
6   Delta_H=886;
7   R=8.31;
8   T0=298;
9
10  beta2=exp(Delta_H/R./T);
11  alpha=alpha0*power(T0./T,2.5);
12
13  max_m=1./(alpha.*beta2);
14
15  h=figure;
16  hold('on');
17  box 'on'
18
19  plot(T,max_m,'r-','LineWidth',3);
20
21
22  set(gca, 'FontSize', 24, 'FontName', 'Times');
23  axis('tight');
24  set(gca,'Position',[.12 .15 .8 .75])
25  xlabel('$T$, K','FontSize',24,'Interpreter','latex');
26  ylabel('$P/P_0$, bar','FontSize',24,'Interpreter','latex');
27
28  str='max_pressure_multilayer';
29  saveas(h,strcat(str,'.png'),'png');
30  saveas(h,strcat(str,'.eps'),'psc2');
```

[1] T. L. Hill, The Journal of Chemical Physics **14**, 263 (1946).

[2] T. Keii, The Journal of Chemical Physics **22**, 1617 (1954).

[3] S. Brunauer, P. H. Emmett, and E. Teller, Journal of the American Chemical Society **60**, 309 (1938).

[4] M. Eddaoudi, D. B. Moler, H. Li, B. Chen, T. M. Reineke, M. O'Keeffe, and O. M. Yaghi, Accounts of Chemical Research **34**, 319 (2001).

[5] A. G. Wong-Foy, A. J. Matzger, and O. M. Yaghi, Journal of the American Chemical Society **128**, 3494 (2006).

[6] S. S. Kaye, A. Dailly, O. M. Yaghi, and J. R. Long, Journal of the American Chemical Society **129**, 14176 (2007).

[7] S. S. Han, D. H. Jung, S.-H. Choi, and J. Heo, ChemPhysChem **14**, 2698 (2013).

[8] L. Landau and E. Lifshitz, *Statistical Physics, Part 1: Volume 5* (Butterworth-Heinemann, 1980).